# A Deep Generative Model for Resting-State EEG Synthesis and Transferable Representation Learning

**Yeganeh Farahzadi[1,2], Morteza Ansarinia[3], Zoltan Kekecs[1]**

[1] Institute of Psychology, Eötvös Loránd University, Budapest, Hungary

[2] Doctoral School of Psychology, Eötvös Loránd University, Budapest, Hungary

[3] Department of Behavioural and Cognitive Sciences, University of Luxembourg, Belval, Luxembourg

## Abstract

Resting-state EEG provides a non-invasive view of spontaneous brain activity, but extracting meaningful patterns is often limited by scarce high-quality data and reliance on manually engineered features. Generative adversarial networks (GANs) can synthesize neural signals and learn transferable representations directly from raw data, a dual capability that remains underexplored in EEG research.

Here, we introduce REST-GAN, a GAN-based framework for resting-state EEG that combines adversarial training with an auxiliary self-supervised reconstruction objective to support signal synthesis and unsupervised feature extraction. Although trained only on raw time-domain signals, without explicit frequency-domain or sensor-topographic supervision, the generated time series reproduced key temporal, spectral, and connectivity properties of real EEG. In band-power feature space, generated samples showed high precision and recall across eyes-open and eyes-closed conditions (EO: 0.91/0.67; EC: 0.87/0.65), while group-average spectral coherence matrices showed low mean absolute differences from real data across frequency bands (~0.01-0.03).

The representations learned by the model's critic transferred to independent resting-state demographic classification tasks, outperforming models trained directly on raw EEG and showing competitive performance relative to a recent EEG foundation model, while requiring substantially less training data and computational resources.

These findings highlight a computationally efficient, architecture-driven strategy in which generative models serve not only as EEG signal generators, but also as unsupervised feature extractors. This approach may support more data-efficient EEG analysis while reducing reliance on manual feature engineering. The implementation code for REST-GAN is available at: https://github.com/Yeganehfrh/REST-GAN.

## 1. Introduction

Generative AI, including techniques such as Generative Adversarial Networks (Goodfellow et al., 2014), Variational Autoencoders (Kingma, 2013), and diffusion models (Ho et al., 2020), is being increasingly recognized as a powerful tool in neuroscience. It has found various applications ranging from synthesizing neural data to also learning reusable representations of neural signals in an unsupervised manner.

One of the challenges in neuroscience is the limited availability of high-quality neural data, as collecting such data is often costly and time-consuming. Generative AI can help address this issue by creating realistic synthetic datasets that capture the statistical properties of real neural recordings. Additionally, in cases where labeled data is sparse or certain conditions are underrepresented, synthetic data can help balance datasets, leading to more robust analyses and improved model performance (Murphy, 2022). The use of generative AI, particularly GANs, for data augmentation in neural datasets has been relatively well studied (Williams et al., 2025). Prior research has demonstrated that augmenting datasets with generated neural signals can enhance classification accuracy in tasks such as motor imagery (Hartmann et al., 2018; Fahimi et al., 2020), emotion recognition (Pan & Zheng, 2021), and detection of epileptic seizure (Pascual et al., 2020).

Beyond data augmentation, generative AI also provides a way to learn meaningful feature representations in an unsupervised manner (Radford, 2015). By learning a latent space that captures meaningful structure in neural signals, GANs can generate interpolated samples that preserve key statistical and physiologically relevant signal properties. These learned representations can be repurposed for supervised tasks, reducing the need for extensive and often labor-intensive feature engineering. This approach is particularly beneficial for neural signal processing, where preprocessing and manual feature extraction can significantly influence final results (Robbins et al., 2020; Botvinik-Nezer et al., 2020). This way, generative unsupervised models trained on large, unlabeled EEG datasets may help address limited labeled data in two complementary ways: by generating additional plausible samples and by learning reusable representations that can be adapted to smaller downstream datasets.

While recent work has introduced novel architectures like criss-cross transformers for unsupervised EEG representation learning (Wang et al., 2024), the use of generative adversarial networks (GANs) for this purpose remains relatively underexplored, with some studies investigating their potential (Liang et al., 2021). One reason may be that GANs in neuroscience have primarily been used for data augmentation, often on relatively small, domain-specific datasets, and frequently to generate extracted features rather than raw time-series data. As a result, they have rarely developed as models that can jointly support realistic neural signal generation and reusable feature learning.

In this study, we extend the use of GANs beyond data augmentation by incorporating larger-scale EEG recordings—particularly resting-state data—to support two linked aims: realistic generation of resting-state EEG and transferable unsupervised representation learning. To align these aims, we combine the adversarial objective with an auxiliary self-supervised reconstruction task. The reconstruction branch encourages the Critic's encoder to retain informative EEG structure rather than relying solely on narrow cues that distinguish real from generated samples, while the adversarial objective keeps the learned features tied to signal realism. This design yields a shared representation space that can support both data synthesis and downstream decoding. We evaluate the transferability of these representations within the resting-state domain using independent downstream datasets.

# 2. Methods

## 2.1. Data

We used the MPI-LEMON dataset (Babayan et al., 2019), a publicly available resource designed to study mind-body-emotion interactions. This dataset includes resting-state EEG (rs-EEG) recordings from 216 participants, utilizing 61 EEG channels during both eye-closed and eye-open experimental blocks. Data from 14 participants were excluded due to missing event information, different sampling rates, mismatching header files, or insufficient data quality (Babayan et al., 2019). Each participant contributed a 16-minute EEG recording, resulting in a total of 54.13 hours of rs-EEG data. During each recording, participants were guided to alternate between 1-minute eyes-open and 1-minute eyes-closed blocks for a total of 16 minutes. These blocks were segmented, then concatenated, and preprocessed identically. Demographic and psychometric metadata (e.g., PANAS, STAI) are also provided in the LEMON dataset but were not used in this study.

### *2.1.1 Preprocessing*

To preserve the natural structure of the EEG data, we kept preprocessing to a minimum (Delorme, 2023). We first downsampled the EEG signals to 128 Hz. Then, following the steps outlined in (Défossez et al., 2023), we applied Baseline correction by subtracting the average value of the first 0.5 seconds from each channel. The data was then normalized using Scikit-Learn's robust scaler, followed by standard scaling. To mitigate the impact of extreme outliers, values exceeding 20 standard deviations were clamped. Data clamping is found to be as effective as more complex artifact rejection methods such as AutoReject (Défossez et al., 2023).

EEG signals were then high-pass filtered at 0.5 Hz to eliminate low-frequency drift. To suppress line noise at 50 Hz, we then applied the Denoising Source Separation (DSS) method (Cheveigné, 2010). The cleaned signals were subsequently segmented into 4-second epochs (512 time points each). Training segments were sampled fully at random: for each segment, a subject and a starting index within that subject's continuous recording were selected uniformly, and a 512-sample window was extracted. Each training step therefore consisted of 128 independently drawn random segments. With 500 training steps per epoch, the model was exposed to 64,000 randomly sampled segments per epoch. Although this procedure does not guarantee exhaustive coverage of the full recordings, it provides highly diverse local temporal contexts during training. Empirically, this stochastic sampling strategy yielded more continuous-looking generated signals when segments were concatenated and reduced visible boundary artifacts, including checkerboard-like patterns.

In this study, we included eight EEG channels (F1, F2, C1, C2, P1, P2, O1, and O2). Selecting 8 out of the 61 EEG channels was primarily for practical reasons, since this subset allowed for faster experimentation and model development. However, the proposed model is designed to be scalable. Specifically, the capacity of each layer—such as the number of filters in the convolutional layers—is proportional to the number of input channels. This allows the model to be adapted to accommodate and process additional EEG channels as needed.

## 2.2. Architecture Details

We implemented a Wasserstein Generative Adversarial Network (Arjovsky et al., 2017) with Gradient Penalty (WGAN-GP) (Gulrajani et al., 2017) to generate multi-channel

EEG data. The model consists of two core components: A generator, responsible for producing synthetic EEG signals, and a discriminator—referred to as a "Critic" in the WGAN framework—which evaluates whether the generated data is real or synthetic.

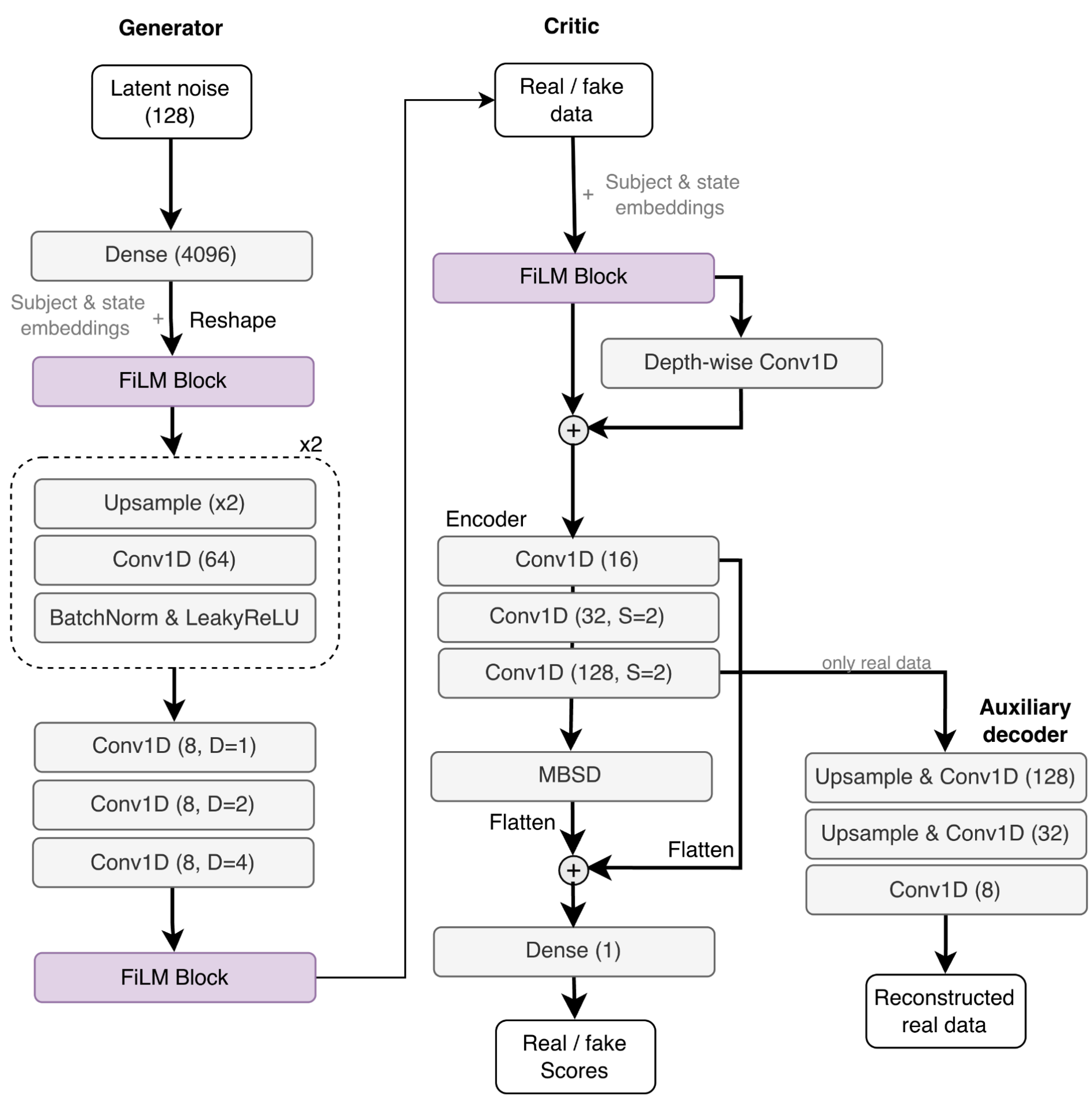

*Figure 1: Model Architecture. The model combines a DCGAN-inspired design with a dilated convolutional block in the Generator and an auxiliary self-supervised regularization in the Critic, intended to improve the modeling of broader temporal dependencies and support EEG representation learning. The Critic further incorporates two inductive biases: (i) a depth-wise convolution acting as a high-pass filter, and (ii) a minibatch standard deviation (MBSD) layer, both discouraging reliance on coarse or trivial adversarial cues. FiLM blocks implement feature-wise linear modulation using subject and condition embeddings. The "+" symbol denotes concatenation of inputs along the feature-map dimension. All convolutional and dense layers use LeakyReLU activations, except for the final Critic output layer, which is linear. In the figure, S denotes stride and D denotes dilation. All upsampling layers double the temporal dimension. Each layer capacity (e.g. number of filters in the convolutional layers) is scalable according to the number of EEG channels.*

Our architecture was inspired by the Deep Convolutional GAN (DCGAN) (Radford, 2015) framework, incorporating a stack of convolutional layers with upsampling layers in the generator and strided convolutional layers in the Critic (Figure 1).

The generator first projects a 128-dimensional latent noise vector into a 4096-dimensional representation, reshapes it to 512 × 8 (time × channels), and applies a FiLM block (described below) to inject subject- and state- specific conditioning (by state, we refer to eye-open or eye-close conditions). Temporal resolution is then progressively increased through two blocks of linear upsampling, Conv1D, Batch Normalization, and LeakyReLU. We used linear interpolation for upsampling, as it has been empirically shown to produce fewer high-frequency artifacts compared to nearest-neighbor upsampling (Hartmann et al., 2018). To improve modeling of broader temporal structure in EEG signals, the upsampled features are further processed by three Conv1D layers with increasing dilation rates (d=1,2,4), following the core idea of Temporal Convolutional Networks (Bai et al., 2018). This dilated block expands the receptive field efficiently, enabling the generator to capture longer-range dependencies without substantially increasing the parameter count. A final FiLM block is applied before output to further align the generated signal with the conditioning variables.

Critic: The critic starts with a FiLM block, followed by a depth-wise Conv1D layer with a fixed kernel, implementing a discrete second derivative independently in each channel. This non-learnable operation acts as a simple high-pass filter that emphasizes sharp local transitions and fast temporal fluctuations before encoding. Its purpose is to make the critic more sensitive to fine-scale temporal artifacts that may be missed when the generator already captures the coarse structure of the signal, thereby encouraging more realistic generation of higher-frequency components.

The resulting representation is then passed to an encoder composed of strided Conv1D layers, which progressively reduces the temporal resolution while extracting higher-level feature maps. The encoded features are subsequently routed to two branches. The adversarial branch outputs the Critic score and begins with an MBSD (mini-batch standard deviation) layer, which appends a scalar summary of variability across the current mini-batch as an additional feature map. This gives the critic access to batch-level diversity cues and improves sensitivity to low-diversity, mode-collapsed generated samples. The auxiliary reconstruction branch is applied only to real data and progressively upsamples the encoded representation to reconstruct the input and predict masked temporal and channel spans. Its loss was weighted by 0.1 to preserve the adversarial focus of the model while still providing a self-supervised regularization signal that encourages more stable and semantically meaningful representations.

All weights in the convolutional and dense layers were initialized using a random normal distribution with a mean of 0 and a standard deviation of 0.02. This initialization has been shown to stabilize GAN training and prevent divergence (Radford, 2015).

FiLM block. We used Feature-Wise Linear Modulation (FiLM; Perez et al., 2018) to condition both the generator and critic on participant identity and recording state. FiLM incorporates auxiliary information by applying feature-wise affine transformations to intermediate activations. In our implementation, each FiLM block receives the current feature representation together with a subject embedding and a state embedding (eye-open vs. eye-closed). These two embeddings are passed through separate linear projections to produce channel-wise scaling (γ) and shifting (β) parameters, which are then combined to modulate the feature maps. Formally, for an input feature map h, the block applies:

$$FiLM(h \mid e_{sub}, e_{state}) = \gamma\left(e_{sub}, e_{state}\right) \odot h + \beta(e_{sub}, e_{state})$$

where γ and β are learned from the subject and state embeddings. To improve training stability, modulation is implemented in a bounded residual form, with γ centered near 1 and both modulation terms down-scaled by 0.1; in addition, the contributions of the subject and state conditionings are each controlled by a learnable scalar gate, allowing the model to attenuate either source of conditioning when less informative. The FiLM projection layers are initialized to zero so that the block begins as an approximate identity mapping at the beginning of training. We used three FiLM blocks in total: two in the generator and one in the critic (Figure 1). This conditioning scheme allows the model to incorporate both subject-specific and state-dependent variability throughout the representation hierarchy in a lightweight and stable manner.

It is important to note that the subject embeddings learned during training are specific to participants in the training set and are therefore not directly available for unseen individuals. However, the FiLM formulation provides a flexible basis for adapting the model to new data. For example, a new subject embedding could be initialized from the population mean of the learned embeddings and optimized using calibration data from the unseen subject while keeping the remaining network weights fixed. Supplementary Section S1 further examines the role of FiLM during inference and generation, showing that the learned subject embeddings contribute meaningfully to the diversity and realism of the generated EEG. We also discuss how interpolation within the learned embedding space can be used to generate samples that are not tied to any single training subject, offering one possible route toward extending generation beyond the discrete subject identities observed during training.

## 2.3. Training

For the training of this model, we used the Wasserstein GAN objective with gradient penalty. Let $G$ denote the generator, which maps a latent noise vector $z \sim p(z)$ to a synthetic EEG sample $G(z)$, and let $D$ denote the critic, which assigns higher scores to real samples than to generated ones. Under the 1-Lipschitz constraint, the critic is trained to approximate the Wasserstein-1 discrepancy between the real data distribution $p_r$ and the generator distribution $p_g$, while the generator is optimized to reduce this discrepancy.

The critic was trained by minimizing

$$L_D = E_{z\sim p(z)}[D(G(z))] - E_{x\sim p_r}[D(x)] + \gamma E_{\hat{x}\sim p_{\hat{x}}}\left[\left(\| \nabla_{\hat{x}} D(\hat{x}) \|_2 - 1\right)^2\right]$$

The first two terms encourage the critic to assign lower scores to generated samples and higher scores to real samples. The final term is the gradient penalty, where $\hat{x} \sim p_{\hat{x}}$ denotes samples drawn uniformly along straight lines between real and generated samples. This penalty promotes the 1-Lipschitz constraint by penalizing deviations of the gradient norm from 1. The gradient penalty coefficient was set to $\lambda = 1$, which provided stable training in our experiments; larger values led to stronger regularization of the critic without improving training behavior.

The generator was trained to maximize the critic's score on generated samples, equivalently by minimizing

$$L_G = E_{z\sim p(z)}[D(G(z))]$$

Both the generator and critic were optimized using the Adam optimizer ($beta_1 = 0.0$, $beta_2 = 0.9$) with an initial learning rate of 0.0002 for the generator and 0.0006 for the critic, with a batch size of 128. The learning rate was reduced progressively using an exponential decay schedule with a decay step of 100,000 and a decay rate of 0.90. Training parameters (e.g. batch size, learning rates) were tuned empirically to ensure stable GAN training and avoid divergence or collapse.

Latent variables z for the generator were sampled from a normal distribution with the mean of 0.0 and standard deviation of 0.1. To ensure the Critic remains well-trained and provides meaningful feedback to the Generator, the Critic was updated three times for each Generator update. Specifically, in each training iteration, three full forward and backward passes were performed on the Critic using both real and generated samples before updating the Generator once.

In the auxiliary reconstruction head, the model is trained to reconstruct masked portions of the input signal from its encoded latent representations (Figure 1). The reconstruction objective is defined as the mean absolute error computed exclusively over the masked elements and normalized by the number of masked positions per sample. To ensure that this auxiliary objective does not overshadow the primary adversarial training, its contribution to the total loss is downweighted by a factor of 0.1.

The model was trained for 420 epochs, corresponding to 210,000 training steps, on an NVIDIA V100 GPU with mixed precision using CUDA 12.7. Model checkpoints were saved every 10 epochs. The checkpoints used for downstream classification and signal generation were selected separately, based on classification performance and visual inspection of generated EEG signals, respectively. The entire model development and training process was implemented using Keras (v3.5.0) with the PyTorch (v2.4.1) backend. We refer to the framework as **REST-GAN** (*Resting-State EEG Synthesis and Transfer-Learning Generative Adversarial Network*). The implementation is available at: https://github.com/Yeganehfrh/REST-GAN

### 2.4. Evaluation of generated EEG realism and diversity across group-average, population, and subject-specific levels

We evaluated the generated EEG at three levels of granularity. At the group level, we compared average waveforms, power spectra, and connectivity. At the population level, we tested whether the set of generated subjects reproduced the distribution, coverage, and internal geometry of the real subjects. At the subject level, we tested whether each generated recording matched its own real counterpart. We additionally tracked how feature-distribution similarity evolved over training. It is important to note that for the metric computation at all levels and the subsequent visualisations, generated time-series segments were first reassembled into the original subject-level format. Thus, instead of evaluating only 4-s training segments, both real and generated data were represented as continuous 8-channel, 8-min recordings for each subject. This subject-level reconstruction allowed us to assess the realism and diversity of the generated EEG in its original continuous format, despite the model being trained on fully randomized short segments.

### *2.4.1 Feature Extraction and subject-level feature spaces*

For each subject, complementary feature sets were extracted from both real and generated EEG.

Spectral features were computed across standard EEG frequency bands: delta, theta, alpha, beta, gamma.

Functional connectivity was quantified using band-limited spectral coherence between all channel pairs. Coherence was estimated using Morlet time–frequency decomposition with seven cycles per frequency. Connectivity was computed separately for consecutive 30-s segments and then averaged across segments, yielding one coherence matrix per subject and frequency band.

Time-domain features included the three Hjorth parameters—activity, mobility, and complexity—which summarize signal variance, frequency composition, and temporal complexity, respectively (Hjorth, 1970).

To characterize second-order temporal structure, each subject's EEG was additionally segmented into overlapping 2-s windows with 1-s step size. For each window, we computed the 8×8 channel covariance matrix. Because covariance matrices are symmetric positive-definite (SPD), each window was represented in the log-Euclidean domain by applying the matrix logarithm and vectorizing the lower triangle including the diagonal, yielding a 36-dimensional log-covariance descriptor. The ordered sequence of these descriptors defined a subject-specific temporal trajectory. This treatment follows the Riemannian framework commonly used for covariance-based EEG analysis (Barachant et al., 2012).

To compare real and generated populations at the subject level, each participant was then represented as a single point in a high-dimensional feature space. For spectral features, each subject was described by a 40-dimensional vector comprising fractional band power across five frequency bands and eight channels. For connectivity features, subjects were represented separately within each frequency band by the 28 unique lower-triangular off-diagonal elements of the corresponding 8×8 coherence matrix. For Riemannian log-covariance features, each subject was represented by a 72-dimensional vector formed by concatenating the mean and standard deviation of the 36-dimensional log-covariance descriptors across windows.

Real and generated datasets therefore formed two subject-level point clouds within each feature space,

$$R = \left\{r_1, \ldots, r_N\right\}, g = \left\{g_1, \ldots, g_N\right\}$$

with N=202 subjects. All features were standardized using the mean and standard deviation of the real data, ensuring that dimensions contributed on a comparable scale while preserving subject-level differences in relative feature magnitude.

### *2.4.2 Group-average similarity*

At the group-average level, real and generated features were averaged across subjects, and we assessed their agreement in three domains: temporal (average waveforms), spectral (power spectral density profiles), and connectivity (inter-channel coherence matrices). Agreement was quantified using Pearson correlation and mean absolute difference for the spectral and connectivity profiles, and signed RMS differences for time-domain amplitude, computed separately for the eyes-open and eyes-closed conditions.

### 2.4.3 Population-level distribution, coverage, and geometry

Moving from group averages to the full population, we asked whether the distribution of generated subjects matched that of the real subjects within each feature space, along three aspects: overall distributional similarity, the balance of fidelity and diversity, and the preservation of inter-subject geometry.

We first quantified overall distributional similarity using the standardized Fréchet Distance (FD) within each subject-level feature space, computed for the complete feature representation at the final training checkpoint. Standardization was used to enable comparisons across features with different numerical scales. Specifically, generated features were transformed using the mean and standard deviation estimated from the corresponding real-data feature distribution, and these same real-data scaling parameters were applied consistently across all model checkpoints.

Because FD provides a single aggregate measure of distributional similarity and does not explicitly distinguish sample fidelity from population diversity, we additionally computed the precision and recall metrics proposed by Kynkäänniemi et al. (2019) in the same subject-level feature spaces. Real and generated manifolds were estimated using local k-nearest-neighbor neighborhoods with distances computed using correlation distance. This choice emphasizes similarity in subject-level feature profiles, rather than absolute feature magnitude. Precision was defined as the proportion of generated samples falling within the real-data manifold and was interpreted as a measure of sample fidelity. Recall was defined as the proportion of real samples falling within the generated-data manifold and was interpreted as a measure of population coverage and diversity.

Finally, we assessed whether the generated cohort preserved the inter-subject geometry of the real cohort using a representational-similarity-style analysis (Kriegeskorte et al., 2008). For each feature space, we computed subject-by-subject representational dissimilarity matrices (RDMs) separately for real and generated data using Euclidean distances between the standardized subject-level feature vectors. This produced two upper-triangular distance sets,

$$D_{RR} = \{d(r_i, r_j)\}_{i<j}, \; D_{GG} = \{d(g_i, g_j)\}_{i<j}$$

describing the internal geometry of the real and generated populations, respectively. Geometry preservation was then quantified as the Spearman correlation between the vectorized upper triangles of these two RDMs. In the present context, this tests whether the generator preserves the relative arrangement of subjects in feature space, rather than merely producing samples that occupy a similar overall region of that space.

As a targeted check on the dominant posterior rhythm, we additionally compared the individual alpha frequency (IAF) between real and generated data. For each subject, IAF was estimated over the posterior channels, and the resulting real and generated IAF distributions were compared using the Wasserstein distance, computed separately for the eyes-open and eyes-closed conditions.

#### *2.4.4. Subject-conditioned fidelity*

Because the generator was explicitly conditioned on subject identity, matching the overall population distribution was not sufficient; it was also necessary to test whether each generated recording reproduced the structure of the specific subject it was conditioned on. For each feature domain, we compared each real subject directly with the generated sample obtained from the same subject identity, yielding paired real–generated similarity estimates. To separate shared population-level structure from finer subject-specific detail, paired similarity was computed both on the raw feature descriptors and after z-normalization across subjects.

### *2.4.5 Training dynamics*

To characterize how real–generated similarity evolved during learning, the standardized FD was additionally computed at regular checkpoints throughout training (every 10 epochs), separately for each electrode region (frontal, central, parietal, occipital). This allowed us to track the learning trajectory of individual feature distributions and to identify features or regions that converged more slowly or non-monotonically.

## 2.5. Downstream classification using Critic encoder representations

To assess whether the representations learned in the intermediate layers of the GAN were transferrable for a new task, we repurposed features learned by the Critic's encoder for two demographic classification tasks: gender and age group classification. Specifically, for each EEG segment, we extracted two encoder representations: an early feature map, h1, obtained after the first convolutional layer, and a deeper feature map, h, obtained after the subsequent strided convolutional blocks. These correspond to the first and final encoder representations shown in Figure 1. Each feature map was mean-pooled across the temporal dimension, and the pooled vectors were concatenated to form a 144-dimensional representation for each segment.

Because the subject embeddings learned during GAN training were tied to the identities present in the source training set, they were not reused for feature extraction on the downstream datasets. Instead, the Critic encoder was provided with the average of the learned training-subject embeddings, yielding a common subject-conditioning vector for all unseen participants and allowing the extracted features to reflect a subject-agnostic representation. Following each train–validation split, encoder features were standardized using the mean and standard deviation estimated from the training fold only, and the same transformation was then applied to the corresponding validation fold. The resulting representations were used as input to a linear classifier with a sigmoid output, equivalent to logistic regression.

For these classification tasks, we used two publicly available EEG datasets from OpenNeuro (Kekecs et al., 2023; Getzmann et al., 2024), both entirely independent of GAN training. The gender-classification dataset comprised eye-closed resting-state EEG from 52 participants (39 female, mean age 24.5), each with approximately 30 minutes of data. To achieve a balanced gender distribution, we resampled the data to include an equal number of male and female participants, resulting in a final subset of 26 participants. The age-classification dataset included eyes-open and eyes-closed resting-state EEG recordings from 608 participants aged between 20 and 70 years, of whom 61.8% were female. For the present study, we used a subset of 280 participants selected to provide balanced representation across both age groups and gender. Participants aged 45 years or younger were assigned to age group 1, whereas participants older than 45 years were assigned to age group 2. The final age-classification sample was balanced across the two age groups. In all cases, only eyes-closed resting-state data were used for the present study.

For train-test splitting, we used StratifiedGroupKFold from Scikit-Learn to preserve class balance while preventing data leakage by ensuring that EEG segments from the same participant did not appear in both the training and test sets (70% train, 30% test). Classifier performance was evaluated using five-fold subject-wise cross-validation, with performance assessed on the held-out validation fold in each split. Balanced accuracy, area under the receiver operating characteristic curve (AUC), and accuracy were used as the primary evaluation metrics. We report the mean and standard deviation of these metrics across the five folds. Models were trained with binary cross-entropy loss and the Adam optimizer (learning rate = 0.001) for up to 1000 epochs, with a batch size of 128, and early stopping was applied based on validation AUC with a patience of 50 epochs.

As baseline, we trained the same classifier directly on minimally processed EEG data from those two held-out datasets as the preprocessing described in the Methods. In this case, each EEG segment was flattened across the time (512) and channel (8) dimensions into a 4096-length vector.

For benchmarking, we compared against CBraMod (Wang et al., 2024), a recently proposed EEG foundation model that has achieved state-of-the-art accuracy on multiple downstream tasks. We trained the same classifier (as described above) on embeddings extracted from the CBraMod model, with no fine-tuning applied to either our model or CBraMod for this downstream task. To ensure a fair comparison, we preprocessed the held-out EEG datasets using the procedure as theirs. Specifically, we selected the same eight EEG channels as in our training set, downsampled the signals to 200 Hz, applied a bandpass filter between 0.3 Hz and 75 Hz, and removed line noise with a 50 Hz notch filter. Then we segmented the data into four-second segments (800 time points; 6400 features) to match the segment length and sample size of the features extracted from our model. Before fitting the classifiers, both CBraMod features and flattened raw EEG inputs were standardized using StandardScaler from scikit-learn, following the same procedure applied to the features extracted from our model.

# 3. Results

## 3.1. The proposed framework generates realistic and diverse multi-channel resting-state EEG

We first assess group-average similarity in the temporal, spectral, and connectivity domains, then evaluate whether the generated cohort reproduces the population-level distribution, coverage, and inter-subject geometry of the real data as well as each subject's own structure, and finally examine how these similarities evolve over training.

### 3.1.1. Group-level temporal, spectral, and connectivity similarity

Figure 2 presents a visual comparison between real and generated EEG signals, highlighting their averaged waveforms, power spectral density, and inter-channel connectivity structure averaged across the subject axis. Despite the model not being explicitly trained on frequency-domain data, and did not have any explicit access to the channel positions and their relative relationship the model has learned to replicate these aspects of the real data, though some systematic bias across channels and frequency bands exists as it's shown in the heatmap and bar summary plots in Figure 2.

In the time domain, signals remain within the same dynamic range demonstrating that the model effectively learned key general temporal properties. Also, comparing generated data in eye-open (EO) and eye-closed (EC) conditions in the time and spectral domain, the model successfully captured the key difference between these two conditions: a more prominent alpha frequency in posterior channels. Summary plots (e, l) showed small signed RMS differences between generated and real signals in both EO and EC. In EO, all regions showed slightly negative values, with the largest underestimation in the occipital region and near-zero differences in parietal, central, and frontal regions. In EC, RMS differences were also slightly negative across regions, with somewhat larger underestimation in central and frontal regions. Overall, the absolute bar magnitudes were small relative to the variability across participants, indicating that time-domain amplitude was well matched on average and that any regional underestimation was modest and not highly consistent across individuals.

Group-averaged power spectral density (PSD) profiles also showed close agreement between real and generated signals in both EO and EC conditions (panels b, i). Across all eight channels, the generated spectra closely followed the canonical shape of the real spectra, including the overall 1/f-like decay and the main spectral peaks. Channel-wise Pearson correlations were uniformly high in both conditions (r=0.96–0.97 in EO; r=0.96–0.98 in EC), indicating strong similarity in the average spectral shape. Mean absolute differences were modest overall. Descriptively, MAD values were slightly higher in EC than in EO, with the largest deviations observed in a subset of EC channels, particularly O2, P2, C1, and F1. The shaded bands further suggest broadly comparable inter-subject variability between real and generated spectra, although the match was not identical in every channel.

Importantly, these correlations quantify similarity between PSD curves averaged across subjects and therefore do not, on their own, establish that the generator captured the full diversity of individual spectral profiles. A model producing highly similar spectra across samples could still achieve strong average PSD correlations while underrepresenting subject-level variability (See Supplementary S1). For this reason, the diversity and coverage of generated spectral features were also examined separately in the population-level geometry analyses.

The PSD difference maps (panels c, j) help localize the remaining spectral mismatch across channels and frequencies. In both EO and EC, the generated data tended to underestimate power at the very lowest frequencies, particularly near the high-pass range. This could reflect slow drifts and low-frequency recording fluctuations that were present in the real data but were not fully reproduced by the generator. A second systematic difference was visible around 50 Hz, where the generated spectra showed lower power than the real data. This suggests that the model did not reproduce the weak residual line-noise component that remained in the real recordings, despite line-noise suppression. In this sense, the generator appeared to capture the physiological spectral profile while attenuating some residual recording-related artifacts.

Beyond the narrow frequency ranges, the heatmaps showed modest deviations, but with a subtle ripple-like pattern of alternating positive and negative differences across frequency. This indicates that the model reproduced the broad spectral shape well, while being less precise in matching the fine-grained local smoothness of the PSD curves. Such small oscillatory deviations may reflect the convolutional/upsampling-based generation process, finite segment length, or Welch-based spectral estimation.

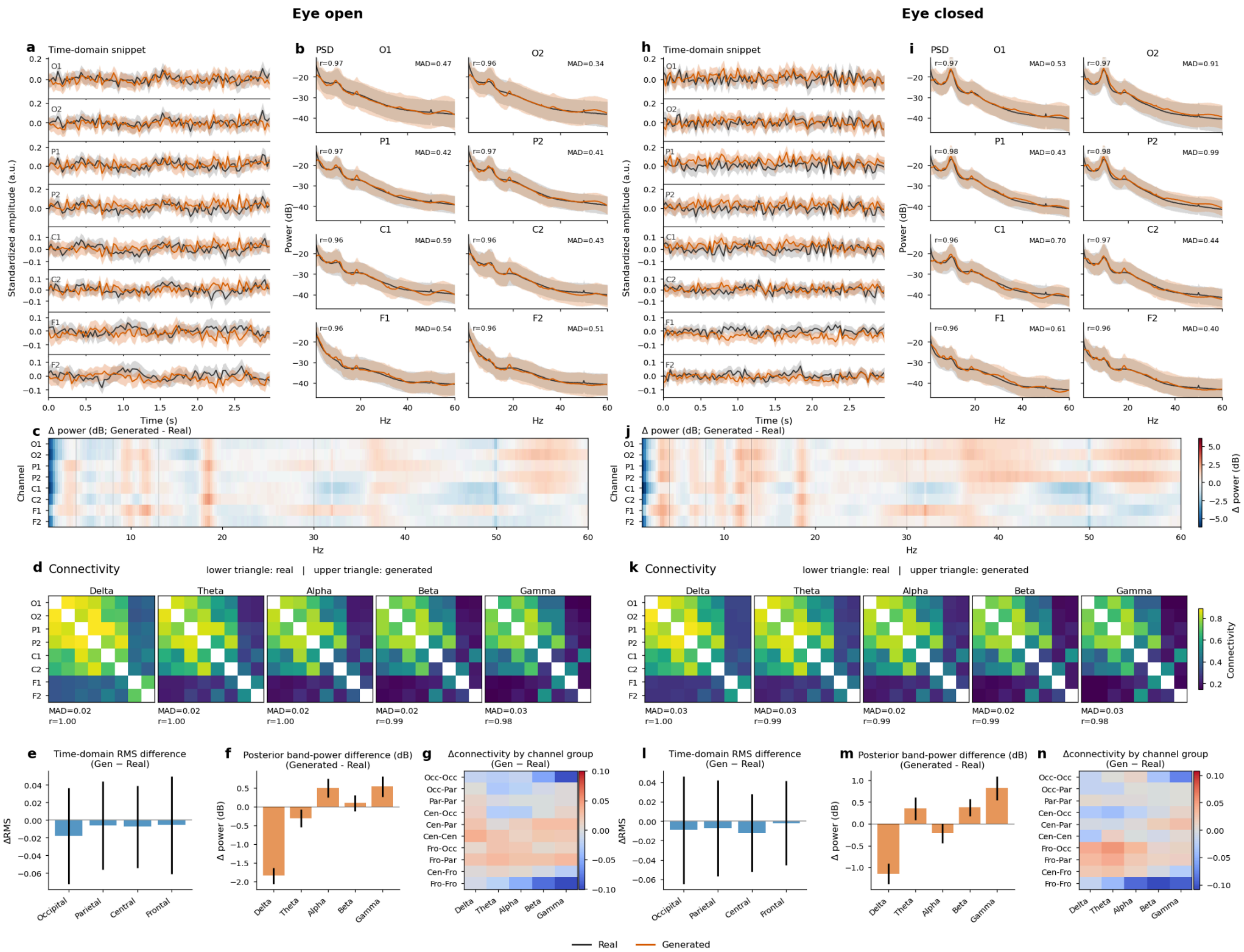

*Figure 2: Visual comparison of generated and real data in the frequency and connectivity domains for eye-open (EO) and eye-closed (EC) conditions, averaged across 202 participants. a, h, Representative 3-s time-domain snippets with shaded variability bands; signals were decimated by a factor of 4 for display. b, i, Power spectral density (PSD) across eight channels; shaded regions indicate standard deviation. For each channel, Pearson correlations (r) and mean absolute differences (MAD) between real and generated PSDs are reported. c, j, Mean PSD difference maps (generated – real) across channels and frequencies. Heatmaps use a zero-centered common scale across EO and EC; vertical lines indicate frequency-band boundaries. d, k, Mean inter-channel coherence matrices for each frequency band, with the lower triangle showing real data and the upper triangle showing generated data. e, l, Signed time-domain RMS differences between generated and real EEG across channel groups. f, m, Posterior band power differences computed over channels P1, P2, O1, and O2. g, n,*

*Coherence across regional channel-pair groups and frequency bands. In line plots, the black line represents the real data, while the orange line shows the generated data.*

To further strengthen the qualitative evaluation, panels d, and k in Figure 2 compare spectral coherence for real and generated EEG across canonical frequency bands. This adds a structure-sensitive perspective beyond the time- and frequency-domain. At the level of the individual-averaged coherence matrices, the generated data closely reproduced the broad topological organization of the real data in both EO and EC, including the overall within- and between-region connectivity structure. Mean absolute differences were uniformly low across bands, approximately 0.01–0.03, and matrix-wise Pearson correlations were near the ceiling (r=0.98–1.00). These results indicate a strong match in the average connectivity organization of each condition. Importantly, because these panels summarize coherence after averaging across subjects, they do not address the preservation of subject-specific connectivity patterns, which is evaluated separately in the population-level analyses.

The regional summary heatmaps in panels g and n further show that this agreement was not spatially uniform. After computing generated-minus-real coherence for each channel pair and averaging these values within regional channel-group combinations, modest region- and band-specific biases became visible. In EO, posterior and posterior–parietal groupings were generally close to zero, whereas slightly larger deviations were observed mainly in connections involving central and frontal regions. In EC, the pattern was somewhat more pronounced: frontal–occipital and frontal–parietal connections tended to show positive generated-minus-real differences, while frontal–frontal connectivity showed negative differences, especially in the higher-frequency bands. Thus, although the mean coherence matrices were highly similar overall, these summary maps indicate small residual biases concentrated in specific region–band combinations. Because each value reflects an average across multiple edge-level differences, values near zero should not be interpreted as perfect agreement at every individual connection, since positive and negative deviations may partly cancel out.

Because these comparisons were performed on the reassembled continuous signals (Section 2.4), the close agreement across time-domain, spectral, and connectivity summaries suggests that, despite being trained on fully randomized segments, the model preserved coherent subject-level EEG structure after stitching. Supplementary Section S2 complements these analyses by projecting real and generated EEG segments into a lower-dimensional UMAP embedding space, enabling qualitative comparison of their overlap and local neighborhood structure, alongside representative matched waveform examples.

#### *3.1.2. Population-level geometry of generated features*

Figure 3 extends the evaluation beyond mean visual similarity by asking whether the generator also reproduced the population geometry and diversity of the real data in Riemannian log-covariance, spectral power, and coherence-based connectivity feature spaces. To this end, real and generated subject populations were compared using Spearman correlations between real and generated subject-level RDMs, precision–recall analysis, posterior individual alpha frequency (IAF), and subject-wise real–generated similarity.

Panels a–c and f–h show the distributions of pairwise inter-subject Euclidean distances derived from the real and generated RDMs for EO and EC, respectively. In both conditions, population geometry was most strongly preserved in the spectral feature space, followed by the log-covariance feature space. This was reflected in the higher Spearman correlations between the vectorized real and generated RDMs for PSD (EO: $\rho$=0.70, EC: $\rho$=0.70) and log-covariance features (EO: $\rho$=0.63, EC: $\rho$=0.54). In contrast, connectivity geometry was reproduced less faithfully, with lower correspondence across coherence bands (EO: $\rho$=0.25–0.42; EC: $\rho$=0.20–0.31). Consistent with the strong PSD-level agreement, posterior IAF distributions showed close real–generated overlap in both EO and EC, with low Wasserstein distances (EO: 0.30; EC: 0.32). Overall, these results indicate that the generator captured the broad manifold structure of spectral variation well, and to a lesser extent that of covariance-based temporal variation, whereas the population geometry of connectivity features was preserved more modestly.

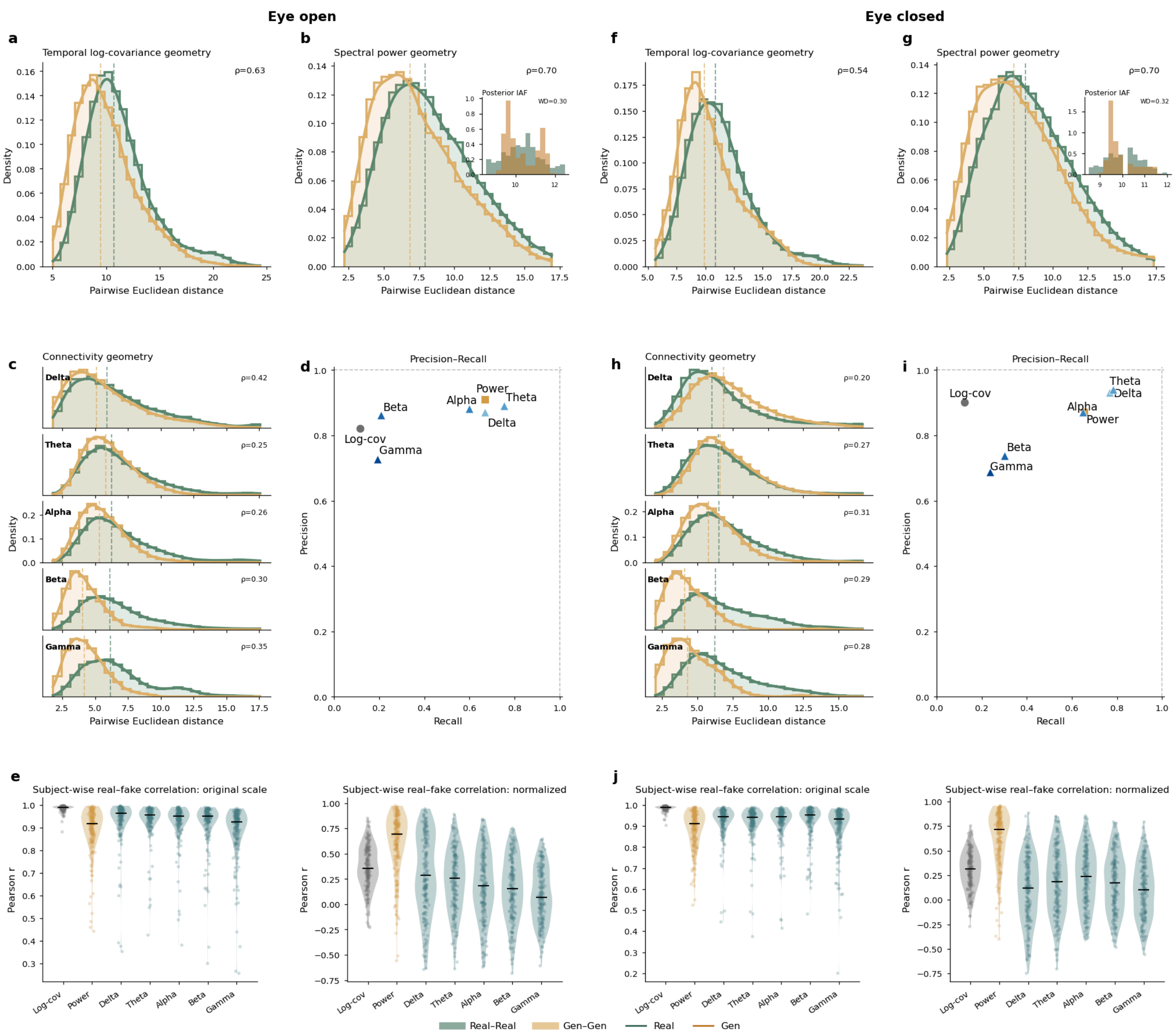

*Figure 3. Population-level geometry, realism, and diversity of real and generated EEG features. a–c, f–h, Distributions of pairwise Euclidean distances between subjects in the real (green) and generated (yellow) populations, shown separately for log-covariance, spectral, and coherence-based connectivity feature spaces. The reported ρ values denote the Spearman correlation between the internal pairwise geometries of the real and generated feature sets. Insets in b and g show the distribution of individual alpha frequency (IAF) estimated over*

*posterior channels, with mismatch between real and generated signals summarized by the Wasserstein distance (WD). d, i, Precision–recall analysis for each feature domain, summarizing fidelity (precision) and diversity (recall) of the generated samples relative to the real-data manifold. e, j, Distributions of paired real–generated subject similarity, quantified as Pearson correlation between matched subject features; the left subpanel shows similarity for raw feature values, and the right subpanel shows similarity after z-transformation, highlighting preservation of both global structure and subject-specific deviations.*

This pattern was broadly consistent with the precision–recall results in panels d and i, which provide a complementary view of sample realism and coverage. In both EO and EC, band-power features showed the most favorable balance of precision and recall, indicating that generated spectral features were both realistic and diverse. Log-covariance features, in contrast, showed the lowest recall in both conditions but relatively high precision, suggesting that the model captured realistic covariance-based temporal structure but did not fully cover the corresponding real-data distribution. Connectivity features were more variable across bands. Theta and alpha coherence showed the strongest precision–recall overall, especially in EC, whereas beta and gamma coherence were less well matched and showed more limited coverage. A modest condition effect was also apparent. EC tended to show slightly stronger precision–recall for band power and for some band-limited connectivity features, particularly theta and delta, consistent with the more structured rhythmic organization of the eyes-closed state.

Panels e and j provide a complementary subject-wise view by correlating each generated subject with its paired real subject in the corresponding feature space. When computed on the raw descriptors, paired similarity was near the ceiling for log-covariance and band power features and remained high for connectivity features in both EC and EO, indicating that the generator captured the dominant structure shared across subjects. After z-normalization across subjects, however, paired similarity decreased markedly—most strongly for connectivity features, more moderately for log-covariance, and least for band power features. This pattern suggests that the model reproduced the population-level template more faithfully than the finer subject-specific deviations that distinguish individuals. In other words, common spectral and log-covariance structure was preserved well, whereas subject-specific "fingerprints," particularly in connectivity, were captured only partially.

Together, these findings suggest that training on randomized short segments was sufficient to capture population-level geometry across feature spaces, although with a clear hierarchy of fidelity: spectral structure was reproduced most consistently, log-covariance structure was captured at the population level but with weaker recall, and connectivity showed band-dependent performance with the greatest loss of individualized detail. These analyses primarily assess coverage within the learned training distribution. Supplementary Section S1 discusses how the FiLM-based conditioning space may also support generation beyond the exact subject embeddings observed during training, for example through interpolation between learned subject representations.

#### *3.1.3. Training dynamics of feature-distribution similarity*

To quantify the similarity between real and generated EEG signals, we computed the standardized Fréchet Distance (FD) on features extracted from both the time and frequency domains. Figure 4 illustrates how FD evolves during training across different electrode regions—frontal, central, parietal, and occipital in EC condition. Table 1 reports the standardized FD values at the final training checkpoint, separately for the EC and EO conditions. To complement the FD-based trajectories, Figure 4b compares the empirical real and generated feature distributions in EC at two checkpoints: early training, epoch 20, and final training, epoch 420. Distributions are shown for each feature across all regions. The corresponding EO distributions are provided in the Supplementary Section S3.

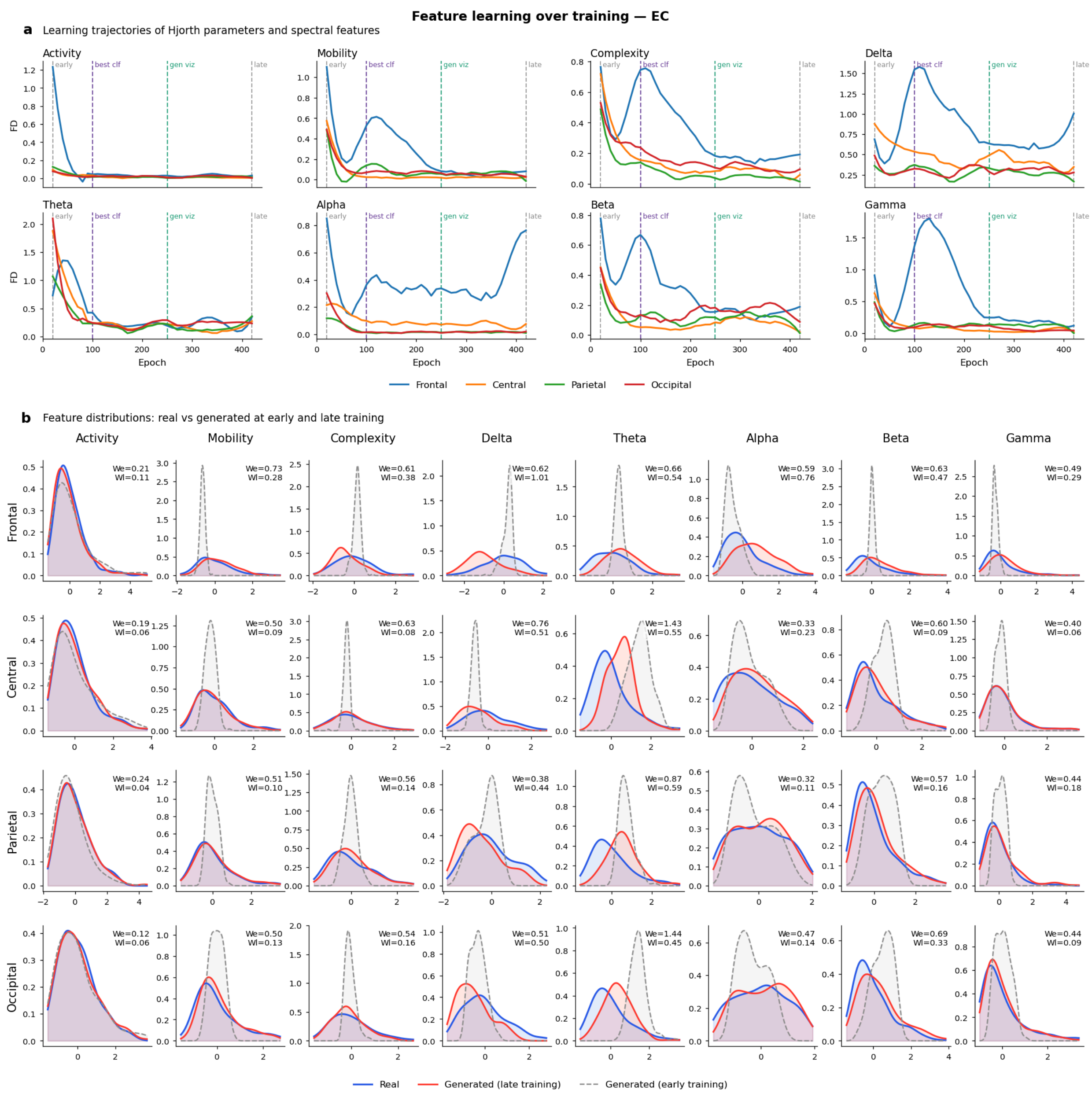

*Figure 4. Feature-distribution similarity during training in the EC condition. a, Standardized Fréchet Distance (FD) between real and generated data across training epochs for Hjorth parameters and spectral EEG features, computed separately for frontal, central, parietal, and occipital regions. Dashed vertical lines indicate key model checkpoints: the checkpoint with the best downstream classification performance (best clf, epoch 100), the checkpoint used for generated-data visualizations (gen viz, epoch 250), and the early and final checkpoints used for the distributional comparisons in panel b: epoch 20 and epoch 420, respectively. b, Regional feature distributions for real data (blue), generated data at the late checkpoint (red), and generated data at the early checkpoint (gray dashed). Values annotated in each subplot indicate the Wasserstein distance between real and generated distributions at the early (We) and late (Wl) checkpoints.*

| Region | Activity | Mobility | Complexity | Delta | Theta | Alpha | Beta | Gamma |
|---|---|---|---|---|---|---|---|---|
| EO | | | | | | | | |
| Frontal | 0.012 | 0.033 | 0.085 | 0.395 | 0.145 | 0.481 | 0.132 | 0.036 |
| Central | 0.011 | 0.025 | 0.052 | 0.246 | 0.390 | 0.041 | 0.054 | 0.031 |
| Parietal | 0.035 | 0.061 | 0.061 | 0.432 | 0.431 | 0.051 | 0.097 | 0.094 |
| Occipital | 0.006 | 0.035 | 0.154 | 0.766 | 0.164 | 0.436 | 0.191 | 0.034 |
| EC | | | | | | | | |
| Frontal | 0.018 | 0.099 | 0.232 | 1.069 | 0.332 | 0.738 | 0.247 | 0.105 |
| Central | 0.004 | 0.029 | 0.053 | 0.329 | 0.300 | 0.081 | 0.012 | 0.023 |
| Parietal | 0.010 | 0.010 | 0.020 | 0.221 | 0.346 | 0.015 | 0.039 | 0.035 |
| Occipital | 0.005 | 0.028 | 0.101 | 0.311 | 0.187 | 0.022 | 0.131 | 0.030 |

Table 1. Standardized *Fréchet Distance (FD) values at epoch 420, for eye-open (EO) and eye-closed (EC) conditions. Lower values indicate closer real–generated feature distributions.*

A breakdown of standardized FD across EEG channels and frequency bands in table 1 reveals a condition-specific regional pattern: In EC, residual mismatches were concentrated mainly in frontal features, whereas in EO the discrepancies were more distributed across regions, with the highest values more often appearing in posterior features. This may reflect differences in the structure of the two resting-state conditions. In EC, the strong and stereotyped posterior alpha rhythm may provide a relatively clear target for the generator, which could explain the low posterior alpha FD and the concentration of residual mismatches in frontal features. In EO, where posterior spectral structure is less dominated by a single rhythm, posterior feature distributions may be more variable and therefore harder to reproduce. Another reason might be that the model may not fully separate EO and EC dynamics and let some generated EO samples partially inherit EC-like posterior structure or vice versa.

Another pattern in Table 1 was the relatively low Hjorth activity FD across regions and conditions, with the strong overlap between real and generated activity distributions in Figure 4b. This pattern suggests that the model captured overall signal variance well, already at the early checkpoint. In other words, gross amplitude structure was learned relatively quickly, although the frontal EO activity trajectory (see Figure S4 in Supplementary Materials) showed less stable convergence before reaching the final value reported in Table 1. This may partly reflect the greater presence of eye-movement or blink-related variance in the eyes-open condition, especially since no targeted ocular artifact removal was applied during preprocessing, making frontal amplitude variability in EO harder to model. In contrast, larger discrepancies were observed for Hjorth complexity, particularly in frontal EC features. Hjorth complexity reflects waveform irregularity and deviations from simple sinusoidal structure; therefore, the higher complexity FD suggests that the model had more difficulty reproducing finer temporal dynamics than overall amplitude. This may reflect the more mixed-frequency and less rhythmically stereotyped nature of frontal signals, where subtle broadband and non-stationary components are likely harder to capture than dominant posterior rhythms.

Consistent with the Hjorth complexity results, alpha power showed a region- and condition-specific pattern. Convergence was strongest for posterior alpha in EC, but weaker for frontal alpha in EC and for frontal/occipital alpha in EO. This suggests that the generator captured the dominant posterior alpha rhythm characteristic of EC resting EEG but reproduced its spatial and condition-specific modulation less accurately. In particular, the residual mismatches indicate that learning the anterior–posterior alpha gradient and the EO–EC change in alpha expression was more challenging than learning the alpha rhythm itself. The alpha distributions support this interpretation. Posterior EC alpha was broad but still reproduced relatively well, indicating that distributional width alone did not account for the mismatch. Instead, the difficulty appeared when alpha activity was less strongly anchored to the dominant EC posterior rhythm, as in frontal EC and parts of EO. Together, these findings suggest that the model learned the dominant spectral signature of EC EEG more robustly than the finer spatial and state-dependent organization of alpha activity.

Among the spectral features, lower-frequency bands, particularly delta, showed weaker convergence across regions. One possible explanation is that delta-band power is more sensitive to residual low-frequency fluctuations, including slow drifts related to sweating, eye movements, blinks, or movement artifacts. Such components are often sparse, non-stationary, condition-dependent, and subject-specific, which may make their distributions harder for the generator to reproduce. In contrast, gamma-band power showed relatively stronger distributional convergence across regions. This may reflect the lower and more stable magnitude of gamma-band power in EEG, where high-frequency power largely follows the descending 1/f-like spectral profile. Interestingly, frontal gamma FD was selectively higher in EC, whereas EO gamma FD was more evenly distributed across regions. This may indicate that in EO, high-frequency variability may be more spatially distributed, leading to a less region-specific gamma mismatch, so that frontal gamma is not as uniquely difficult relative to the other regions. This interpretation remains tentative, as gamma-band EEG is also particularly sensitive to residual artifacts and should not be overinterpreted as purely neural activity.

Beyond the final-epoch regional differences, the training trajectories in figure 4 reveal a prominent transient increase in frontal FD during the intermediate phase of training. This effect was visible across several frontal features, including mobility, complexity, delta, beta, and gamma, with most trajectories peaking around epoch 100 before gradually declining. This pattern suggests that frontal temporal and broadband spectral features were not learned monotonically but instead underwent a period of increased mismatch while the adversarial dynamics were still evolving. One possible interpretation is that, during this phase, the generator had not yet stabilized its approximation of frontal feature distributions, while the critic remained highly sensitive to frontal discrepancies between real and generated signals. Interestingly, this intermediate peak coincided with the checkpoint that yielded the strongest downstream age and gender classification performance from the critic's encoded features. This temporal correspondence suggests that the most useful critic representations may emerge before the generator reaches its best feature-level fidelity, when the critic is still strongly discriminating real from generated samples and may preserve subject-relevant frontal information. However, this relationship should be interpreted cautiously: higher FD reflects poorer generator-feature matching and does not, by itself, imply better representation quality.

## 3.2. The Critic learns transferable EEG representation for downstream demographic prediction

To evaluate the transferability of the Critic's encoder learned representations, we tested its performance on two downstream demographic classification tasks: gender and age-group decoding. We compared our features with flattened raw EEG segments and features extracted from CBraMod. As shown in Table 2, our representation performed comparably to CBraMod on gender classification, yielding slightly higher accuracy (0.683 vs. 0.664) but lower AUROC (0.722 vs. 0.752). For age-group classification, CBraMod showed stronger performance, reaching an accuracy of 0.687 and AUROC of 0.756, compared with 0.632 and 0.687 for our model. Both learned representations clearly outperformed flattened raw EEG, indicating that downstream linear decodability was not simply driven by input dimensionality.

| Task | Dataset size | Feature source | Feature dimension | Val. accuracy | Val. balanced accuracy | Val. AUROC |
|---|---|---|---|---|---|---|
| Gender | 26 subjects / 10,608 segments | Flattened raw EEG | 4096 | 0.503 ± 0.012 | 0.503 ± 0.009 | 0.506 ± 0.008 |
| Gender | 26 subjects / 10,608 segments | REST-GAN critic (ours) | 144 | 0.683 ± 0.100 | 0.683 ± 0.105 | 0.722 ± 0.153 |
| Gender | 26 subjects / 10,608 segments | CBraMod | 6400 | 0.664 ± 0.067 | 0.679 ± 0.069 | 0.752 ± 0.079 |
| Age group | 280 subjects / 12,600 segments | Flattened raw EEG | 4096 | 0.516 ± 0.010 | 0.516 ± 0.010 | 0.520 ± 0.011 |
| Age group | 280 subjects / 12,600 segments | REST-GAN critic (ours) | 144 | 0.632 ± 0.020 | 0.632 ± 0.020 | 0.687 ± 0.022 |

| Age group | 280 subjects / 12,600 segments | CBraMod | 6400 | 0.687 ± 0.038 | 0.687 ± 0.038 | 0.756 ± 0.047 |
|---|---|---|---|---|---|---|

*Table 2. Logistic regression classification performance across downstream demographic tasks. All models were evaluated using the same deterministic StratifiedGroupKFold splits and the same logistic-regression classifier. Values show mean ± SD across five folds. Feature dimension corresponds to the length of each input sample's shape before classification. The top-performing result is bolded.*

This comparison should also be interpreted in light of the substantial difference in pretraining scale. As summarized in Table 3, CBraMod was trained on approximately 333× more EEG duration, 523× more timepoints, and about 120× more GPU-hours than our model, while producing a 44× higher-dimensional feature representation. CBraMod was also slower during downstream use: feature extraction took approximately 10.2× longer for the gender dataset and 12.7× longer for the age-group dataset, and training the linear classifier required ~7–8 ms/step compared with ~3–4 ms/step for our lower-dimensional features. Together, these results suggest that the Critic learns compact and transferable EEG representations that remain competitive under far more limited pretraining and downstream computational demands.

| Model | Pretraining scale | Pretraining resources | Downstream feature dim | Feature extraction time | Classifier training |
|---|---|---|---|---|---|
| REST-GAN | ~27 h EEG; ~12.4M timepoints; 8 channels; downstream checkpoint: 50k steps (100 epochs) | 1× NVIDIA Tesla V100 16GB; ~4 h to epoch-100 checkpoint | 144 | Gender: 4.1 s; Age: 1.8 s | 3–4 ms/step |
| CBraMod | ~9,000 h EEG; ~6.48B timepoints; 32 channels; 40 epochs | 4× NVIDIA RTX A5000; ~5 days | 6400 | Gender: 41.8 s; Age: 22.8 s | 7–8 ms/step |

*Table 3. Pretraining and downstream classification computational cost. Feature-extraction times were measured after a hot start. For downstream classification, both models were used as frozen feature extractors and only a single sigmoid classifier was trained. Classifier training times were measured on the same hardware setup used for the downstream analyses. CBraMod features were extracted from the same eight EEG channels used by our model, using CBraMod-specific preprocessing.*

Taken together, these results suggest that our model learns a compact and computationally efficient EEG representation, highlighting the promise of GAN-based unsupervised representation learning, as a resource-efficient alternative to large, data-intensive foundation models and a potential direction to reducing reliance on manual feature engineering.

## 4. Discussion and Conclusion

In this study, we implemented a Wasserstein Generative Adversarial Network with Gradient Penalty (WGAN-GP) augmented with an auxiliary self-supervised reconstruction task to generate multi-channel resting-state EEG and learn unsupervised representations within the same training process. The realism and diversity of the synthesized signals were evaluated at three levels of granularity: at the group level, we compared reconstructed continuous EEG in the temporal, spectral, and connectivity domains; at the population level, we assessed the distribution, fidelity, coverage, and inter-subject geometry of the generated cohort; and at the subject level, we examined how closely each generated recording matched its own real counterpart. We additionally tracked how feature-distribution similarity evolved over training.

Overall, the results demonstrate that the model learned key, temporal, spectral, and inter-channel characteristics of EEG signals. Generated data closely reproduced the average spectral and connectivity structure of real EEG, while also preserving feature-level diversity. Performance, however, varied across feature spaces: spectral properties were reproduced most consistently, log-covariance descriptors showed high realism but more limited coverage of the full real-data distribution, and connectivity features retained broad topological organization but showed weaker preservation of fine-grained, subject-specific structure. Notably, these properties emerged even though the model was trained directly on raw time-domain segments, without explicit supervision on spectral or connectivity features. This suggests that the combined adversarial and self-supervised objectives were effective to drive the model toward learning physiologically relevant sensor-level EEG structure beyond local waveform appearance alone.

The standardized Fréchet Distance (FD) trajectories further provided a window into the model's learning dynamics across features, conditions, and region groups. A clear pattern was that gross amplitude structure, reflected by Hjorth activity, converged relatively quickly across both EO and EC conditions. By contrast, more complex temporal properties, particularly Hjorth complexity, remained harder to match, with the largest residual discrepancies observed in the frontal channels. This suggests that the model learned overall variance more readily than finer waveform irregularity and non-sinusoidal temporal structure. A second pattern emerged in the alpha range. The dominant posterior alpha rhythm in EC was reproduced relatively well, whereas convergence was weaker for frontal alpha in EC and for posterior alpha in EO. This indicates that the generator captured the salient spectral signature of eyes-closed EEG more robustly than its finer spatial and state-dependent modulation. In other words, learning the presence of a strong posterior alpha rhythm appeared easier than learning the full anterior–posterior and EO–EC alpha gradients.

The spectral FD trajectories also showed that lower-frequency features, particularly delta-band power, remained more difficult to reproduce than higher-frequency gamma-band power. This likely reflects the greater sensitivity of delta power to slow, non-stationary fluctuations and residual artifact-related variability, rather than a simple frequency-dependent learning preference. Gamma-band convergence, in contrast, was generally stronger, although its interpretation remains cautious because high-frequency EEG is also vulnerable to residual non-neural contributions. Together, these results suggest that the model's learning difficulty was shaped less by frequency alone than by the stability, stationarity, and condition specificity of the underlying feature distributions.

Finally, the checkpoint that yielded the strongest downstream demographic classification performance did not coincide with the checkpoint that showed the best generative fidelity. Instead, the best classification checkpoint occurred during an intermediate phase of training, when several frontal feature FD trajectories showed transient increases before later declining. This suggests that the critic's most transferable representations may emerge before the generator reaches its strongest feature-level realism, potentially while the critic remains especially sensitive to discriminative differences between real and generated EEG.

Importantly, these interpretations remain observational and the present analyses were not designed to formally test specific mechanisms of GAN learning. Controlled follow-up experiments will be needed to determine which architectural choices, training dynamics, and signal properties directly shape the emergence of generative realism and transferable EEG representations.

Compared with prior work, the main shift in emphasis is from augmentation toward representation learning.  GANs in EEG have been used predominantly for data augmentation in task-specific BCI settings (e.g., Fahimi et al., 2020), frequently generating single-channel signals (e.g., Hartmann et al., 2018) or extracted features rather than raw time series (e.g., Luo et al., 2020). Building on these foundational efforts, we instead target task-free, resting-state EEG — still comparatively underexplored, with few examples such as Mahey et al. (2023) — and treat the trained model not only as a generator but as a source of reusable, unsupervised representations. The emphasis is therefore less on the fact of synthesis than on what the learned representations enable downstream, at a fraction of the data and compute of large foundation models.

This focus on resting-state data is deliberate: recorded without any task, it offers a task-independent  substrate for learning the intrinsic spectral, spatial, and inter-individual structure of neural signals. Such representations may provide a useful foundation for downstream analyses and could potentially support transfer to task-specific applications. This possibility is motivated by prior neuroimaging work showing that resting-state activity reflects intrinsic brain organization corresponding to task-evoked networks (Smith et al., 2009), constrains task activations through activity flow mechanisms (Cole et al., 2016), and predicts individual differences in task-evoked responses (Tavor et al., 2016). However, the present findings do not establish direct generalization to task-evoked activity, and the utility of these representations for task-specific EEG remains an important direction for future work.

Another contribution of the present study is the use of a large resting-state EEG dataset spanning hundreds of participants, combined with FiLM-based conditioning to account for inter-individual variability. This approach represents participants within a continuous embedding space and uses these learned embeddings to modulate intermediate network features during training and generation. Our findings suggest that FiLM conditioning contributed to preserving subject-level variability and supporting diversity in the synthesized signals, while also providing a flexible route for generating samples beyond the discrete subject identities seen during training, for example through interpolation within the learned subject embedding space. More broadly, this design directly addresses a well-known challenge in EEG modeling: the substantial variability in signal structure across individuals (Saha, S., & Baumert, M., 2020; Kamrud, A. et al., 2021).

Finally, EEG generation studies often rely on improved classification accuracy after data augmentation as the primary proxy for data quality. While this is a useful criterion, it does not fully capture signal realism or diversity, and broader evaluations are often overlooked (see Habashi et al., 2023, for a review). More generally, defining "realistic EEG" is inherently metric-dependent, and no single measure can capture all physiologically relevant properties of resting-state signals. In the present study, we evaluated whether generated signals reproduced key temporal, spectral, and connectivity properties of real EEG at the group, population, and subject levels. However, these metrics do not exhaustively characterize EEG realism. For example, the current evaluation does not explicitly assess cross-frequency interactions, microstate dynamics, or source-level physiological plausibility. More importantly, similarity at the sensor level does not guarantee that generated signals arise from the same underlying neural generators, because the EEG inverse problem is non-unique and multiple source configurations can produce similar scalp potentials (Grech et al., 2008; Michel & Brunet, 2019). Therefore, our results should be interpreted as evidence of sensor-level realism across several resting-state EEG properties, but not full neurophysiological equivalence between real and generated signals. Moving forward, evaluation strategies based on deep feature embeddings, latent-space geometry, or source-level modeling may provide a more faithful assessment of generative quality and help probe the plausibility of the underlying neural structure reflected in generated EEG signals.

At the same time, the downstream transfer results suggest that the model learned more than superficial sensor-level statistics. Without fine-tuning the Critic, we extracted its intermediate representations and trained only a simple linear classifier for gender and age-group classification on two independent resting-state datasets. In both tasks, Critic-derived features clearly outperformed flattened raw EEG, indicating that the learned feature space contains structured, linearly decodable information rather than merely preserving low-level signal variance. Despite their compact 144-dimensional form, these features achieved performance comparable to CBraMod for gender classification, although CBraMod performed better for age-group classification. This difference is informative rather than unexpected, given that CBraMod was pretrained on substantially larger datasets, with broader channel coverage and far greater computational resources. Together, these findings suggest that the Critic learns compact and transferable EEG representations, while larger foundation models may capture a broader range of demographic information.

Also, to keep the downstream evaluation conservative and directly comparable with CBraMod, we did not adapt subject-specific embeddings for the unseen participants. Instead, we used the mean subject embedding learned during pretraining as a universal conditioning vector for all participants in the downstream datasets. The fact that the extracted features remained predictive under this setting further suggests that the Critic learned representations that extend beyond the original training subjects. Future work could build on this by learning subject-specific embeddings for newly enrolled participants from a small amount of calibration data, or by training an auxiliary encoder to infer subject embeddings directly from short EEG segments.

These results also highlight an often-overlooked aspect of GAN-based approaches: the Critic is not only a training signal for the generator, but also a potential representation learner. A representationally weak or collapsed Critic limits the feedback available to the generator, whereas a Critic that captures meaningful temporal, spectral, and inter-channel structure can provide richer guidance during adversarial learning. In this sense, improving Critic representations may benefit both downstream feature extraction and signal generation.

Despite the promising results, our approach has limitations. First, the training data was derived from a single dataset (LEMON) and limited to eight selected EEG channels, which may restrict the model's ability to generalize to datasets with different recording conditions or additional EEG channels distributed across the scalp. Second, although the learned representations showed transferability across resting-state downstream tasks, the present study does not establish whether these representations generalize to task-evoked EEG activity, such as motor imagery, emotion recognition, or other event-related paradigms that are better studied in this GAN-based EEG generation line of research.

Building upon the current work, several directions can be pursued in future research; first expanding the training data by incorporating larger, and more diverse EEG datasets, including task-related and clinical recordings, would improve the model's generalizability and enable direct evaluation across different EEG domains. Future work could also examine broader applications of the learned representations, such as cognitive state decoding and neurological disorder detection. In addition, improving model interpretability remains an important next step. Analyses of the GAN's latent space, the contribution of individual EEG channels, and the features extracted by the Critic could help clarify what aspects of the signal are being captured by the model.

Another promising direction is to use generative EEG models for controlled method validation with known ground truth, rather than treating synthetic EEG as a substitute for real EEG. For example, relatively clean EEG segments could be contaminated with defined artifacts components, such as ocular activity, and a generative denoising model could then be tested on whether it can suppress the added artifact while recovering the original clean signal. The potential advantage of a GAN-based approach is that the adversarial objective can encourage the restored signal to match the broader distributional properties of real clean EEG, rather than relying only on pointwise reconstruction losses such as MSE or MAE. This would provide a more controlled way to test artifact-removal pipelines, which is difficult with purely real EEG because the underlying clean neural signal is usually unknown. This direction also aligns with existing GAN-based approaches to EEG denoising (Sawangjai et al., 2022; Dong et al., 2023), while extending them toward controlled artifact manipulation and validation.

Taken together, our findings underscore the dual utility of GAN architecture: the generator learns to produce realistic EEG signals, while the Critic functions as an efficient unsupervised feature extractor that supports downstream classification. Although the comparison with CBraMod was limited to two demographic tasks, it suggests that GAN-based models can offer a resource-efficient alternative for EEG representation learning, particularly when large-scale foundation-model pretraining is not feasible. As the field increasingly moves toward general-purpose neural foundation models and generative AI, such dual-purpose GAN architectures may offer a data-efficient route in neuroscience, bridging high-fidelity signal synthesis with reusable EEG representation learning.

## Acknowledgment

This research was supported by the Hungarian National Research, Development and Innovation Office (NKFIH, Grant No.: FK 132248).

## References

Arjovsky, M., Chintala, S., & Bottou, L. (2017). Wasserstein generative adversarial networks. International Conference on Machine Learning, 214–223.

Barachant, A., Bonnet, S., Congedo, M., & Jutten, C. (2012). Multiclass brain–computer interface classification by Riemannian geometry. IEEE Transactions on Biomedical Engineering, 59(4), 920-928.

Babayan, A., Erbey, M., Kumral, D., Reinelt, J. D., Reiter, A. M., Röbbig, J., Schaare, H. L., Uhlig, M., Anwander, A., Bazin, P.-L., et al. (2019). A mind-brain-body dataset of MRI, EEG, cognition, emotion, and peripheral physiology in young and old adults. Scientific Data, 6(1), 1–21.

Bai, S., Kolter, J. Z., & Koltun, V. (2018). An empirical evaluation of generic convolutional and recurrent networks for sequence modeling. arXiv preprint arXiv:1803.01271.

Botvinik-Nezer, R., Holzmeister, F., Camerer, C. F., Dreber, A., Huber, J., Johannesson, M., Kirchler, M., Iwanir, R., Mumford, J. A., Adcock, R. A., et al. (2020). Variability in the analysis of a single neuroimaging dataset by many teams. Nature, 582(7810), 84–88.

Cheveigné, A. de. (2010). Time-shift denoising source separation. Journal of Neuroscience Methods, 189(1), 113–120.

Cole, M. W., Ito, T., Bassett, D. S., & Schultz, D. H. (2016). Activity flow over resting-state networks shapes cognitive task activations. Nature neuroscience, 19(12), 1718-1726.

Défossez, A., Caucheteux, C., Rapin, J., Kabeli, O., & King, J.-R. (2023). Decoding speech perception from non-invasive brain recordings. Nature Machine Intelligence, 5(10), 1097–1107.

Delorme, A. (2023). EEG is better left alone. Scientific Reports, 13(1), 2372.

Dong, Y., Tang, X., Li, Q., Wang, Y., Jiang, N., Tian, L., ... & Fang, P. (2023). An approach for EEG denoising based on Wasserstein generative adversarial network. IEEE Transactions on Neural Systems and Rehabilitation Engineering, 31, 3524-3534.

Fahimi, F., Dosen, S., Ang, K. K., Mrachacz-Kersting, N., & Guan, C. (2020). Generative adversarial networks-based data augmentation for brain–computer interface. IEEE Transactions on Neural Networks and Learning Systems, 32(9), 4039–4051.

Farahzadi, Y., Alldredge, C., & Kekecs, Z. (2024). Gamma power and beta envelope correlation are potential neural predictors of deep hypnosis. Scientific Reports, 14(1), 6329.

Getzmann, S., Gajewski, P. D., Schneider, D., & Wascher, E. (2024). Resting-state EEG data before and after cognitive activity across the adult lifespan and a 5-year follow-up [Dataset]. OpenNeuro. https://doi.org/10.18112/openneuro.ds005385.v1.0.0

Goodfellow, I., Pouget-Abadie, J., Mirza, M., Xu, B., Warde-Farley, D., Ozair, S., Courville, A., & Bengio, Y. (2014). Generative adversarial nets. Advances in Neural Information Processing Systems, 27.

Grech, R., Cassar, T., Muscat, J., Camilleri, K. P., Fabri, S. G., Zervakis, M., ... & Vanrumste, B. (2008). Review on solving the inverse problem in EEG source analysis. Journal of neuroengineering and rehabilitation, 5(1), 25.

Gulrajani, I., Ahmed, F., Arjovsky, M., Dumoulin, V., & Courville, A. C. (2017). Improved training of wasserstein gans. Advances in Neural Information Processing Systems, 30.

Habashi, A. G., Azab, A. M., Eldawlatly, S., & Aly, G. M. (2023). Generative adversarial networks in EEG analysis: An overview. Journal of Neuroengineering and Rehabilitation, 20(1), 40.

Hartmann, K. G., Schirrmeister, R. T., & Ball, T. (2018). EEG-GAN: Generative adversarial networks for electroencephalograhic (EEG) brain signals. arXiv Preprint arXiv:1806.01875.

Hjorth, B. (1970). EEG analysis based on time domain properties. Electroencephalography and Clinical Neurophysiology, 29(3), 306–310.

Ho, J., Jain, A., & Abbeel, P. (2020). Denoising diffusion probabilistic models. Advances in Neural Information Processing Systems, 33, 6840–6851.

Kamrud, A., Borghetti, B., & Schubert Kabban, C. (2021). The effects of individual differences, non-stationarity, and the importance of data partitioning decisions for training and testing of EEG cross-participant models. Sensors, 21(9), 3225

Kekecs, Z., Girán, K., Vizkievicz, V., Lutoskin, A., & Farahzadi, Y. (2023). *The effects of sham hypnosis techniques* [Dataset]. OpenNeuro. https://doi.org/10.18112/openneuro.ds004572.v1.3.0

Kingma, D. P. (2013). Auto-encoding variational bayes. arXiv Preprint arXiv:1312.6114.

Kriegeskorte, N., Mur, M., & Bandettini, P. A. (2008). Representational similarity analysis-connecting the branches of systems neuroscience. Frontiers in systems neuroscience, 2, 249.

Kynkäänniemi, T., Karras, T., Laine, S., Lehtinen, J., & Aila, T. (2019). Improved precision and recall metric for assessing generative models. Advances in neural information processing systems, 32.

Liang, Z., Zhou, R., Zhang, L., Li, L., Huang, G., Zhang, Z., & Ishii, S. (2021). EEGFuseNet: Hybrid unsupervised deep feature characterization and fusion for high-dimensional EEG with an application to emotion recognition. IEEE Transactions on Neural Systems and Rehabilitation Engineering, 29, 1913–1925.

Luo, Y., Zhu, L.-Z., Wan, Z.-Y., & Lu, B.-L. (2020). Data augmentation for enhancing EEG-based emotion recognition with deep generative models. Journal of Neural Engineering, 17(5), 056021.

Mahey, P., Toussi, N., Purnomu, G., & Herdman, A. T. (2023). Generative adversarial network (GAN) for simulating electroencephalography. Brain Topography, 36(5), 661-670.

Michel, C. M., & Brunet, D. (2019). EEG source imaging: a practical review of the analysis steps. Frontiers in neurology, 10, 325.

Murphy, K. P. (2022). Probabilistic machine learning: An introduction. MIT press.

Pan, B., & Zheng, W. (2021). Emotion recognition based on EEG using generative adversarial nets and convolutional neural network. Computational and Mathematical Methods in Medicine, 2021(1), 2520394.

Pascual, D., Amirshahi, A., Aminifar, A., Atienza, D., Ryvlin, P., & Wattenhofer, R. (2020). Epilepsygan: Synthetic epileptic brain activities with privacy preservation. IEEE Transactions on Biomedical Engineering, 68(8), 2435–2446.

Perez, E., Strub, F., De Vries, H., Dumoulin, V., & Courville, A. (2018, April). Film: Visual reasoning with a general conditioning layer. In Proceedings of the AAAI conference on artificial intelligence (Vol. 32, No. 1).

Radford, A. (2015). Unsupervised representation learning with deep convolutional generative adversarial networks. arXiv Preprint arXiv:1511.06434.

Robbins, K. A., Touryan, J., Mullen, T., Kothe, C., & Bigdely-Shamlo, N. (2020). How sensitive are EEG results to preprocessing methods: A benchmarking study. IEEE Transactions on Neural Systems and Rehabilitation Engineering, 28(5), 1081–1090.

Saha, S., & Baumert, M. (2020). Intra-and inter-subject variability in EEG-based sensorimotor brain computer interface: a review. Frontiers in computational neuroscience, 13, 87.

Smith, S. M., Fox, P. T., Miller, K. L., Glahn, D. C., Fox, P. M., Mackay, C. E., ... & Beckmann, C. F. (2009). Correspondence of the brain's functional architecture during activation and rest. Proceedings of the national academy of sciences, 106(31), 13040-13045.

Sawangjai, P., Trakulruangroj, M., Boonnag, C., Piriyajitakonkij, M., Tripathy, R. K., Sudhawiyangkul, T., & Wilaiprasitporn, T. (2021). EEGANet: Removal of ocular artifacts from the EEG signal using generative adversarial networks. IEEE Journal of Biomedical and Health Informatics, 26(10), 4913-4924.

Tavor, I., Jones, O. P., Mars, R. B., Smith, S. M., Behrens, T. E., & Jbabdi, S. (2016). Task-free MRI predicts individual differences in brain activity during task performance. Science, 352(6282), 216-220.

Wang, J., Zhao, S., Luo, Z., Zhou, Y., Jiang, H., Li, S., Li, T., & Pan, G. (2024). CBraMod: A criss-cross brain foundation model for EEG decoding. arXiv Preprint arXiv:2412.07236.

Williams, C. C., Weinhardt, D., Hewson, J., Plomecka, M. B., Langer, N., & Musslick, S. (2025). EEG-GAN: A Generative EEG Augmentation Toolkit for Enhancing Neural Classification. bioRxiv, 2025-06.

# Supplementary Materials

## S1: Effects of subject FiLM conditioning on generated EEG

To better understand the role of the subject-conditioning layers in shaping the realism and diversity of the generated signals, we performed an inference-time sensitivity analysis. Using the same trained generator, we generated samples under two modified settings: first, by replacing all subject-specific embeddings with the mean of the learned subject embeddings (“mean subject embedding”), and second, by bypassing the subject-conditioning in the FiLM layers during generation while retaining the eye-condition modulation (“no subject modulation”). The results are shown in Figure S1.

In both settings, recall dropped to zero, indicating a marked reduction in the diversity of the generated samples once subject-specific modulation was removed or collapsed to a single population-level embedding. Precision, however, remained very high for temporal log-covariance features in the mean-embedding condition. This pattern suggests that the generator produced conservative, average-like signals that still fell within the real-data manifold but covered only a narrow part of it. In other words, the generated samples were still locally realistic but no longer captured the broader variability present in the real data.

A similar pattern was visible in the PSD results. The average generated PSDs remained highly correlated with the real PSDs, but the generated spectra showed reduced variability compared with the real data. This highlights a limitation of relying only on averaged similarity measures: high correlation at the population level can mask a loss of sample-level diversity. Together, these findings suggest that the FiLM-based subject modulation contributes not only to spectral fidelity, but also to preserving inter-individual variability in the generated EEG.

It is important to note that this analysis was performed by modifying the conditioning inputs at inference time while keeping all trained weights fixed. Therefore, it should be interpreted as a sensitivity analysis of the trained generator rather than a full architectural ablation. A full ablation, in which the model is retrained without the subject-conditioning layers, would be needed to isolate the complete contribution of FiLM conditioning to training dynamics and generation quality.

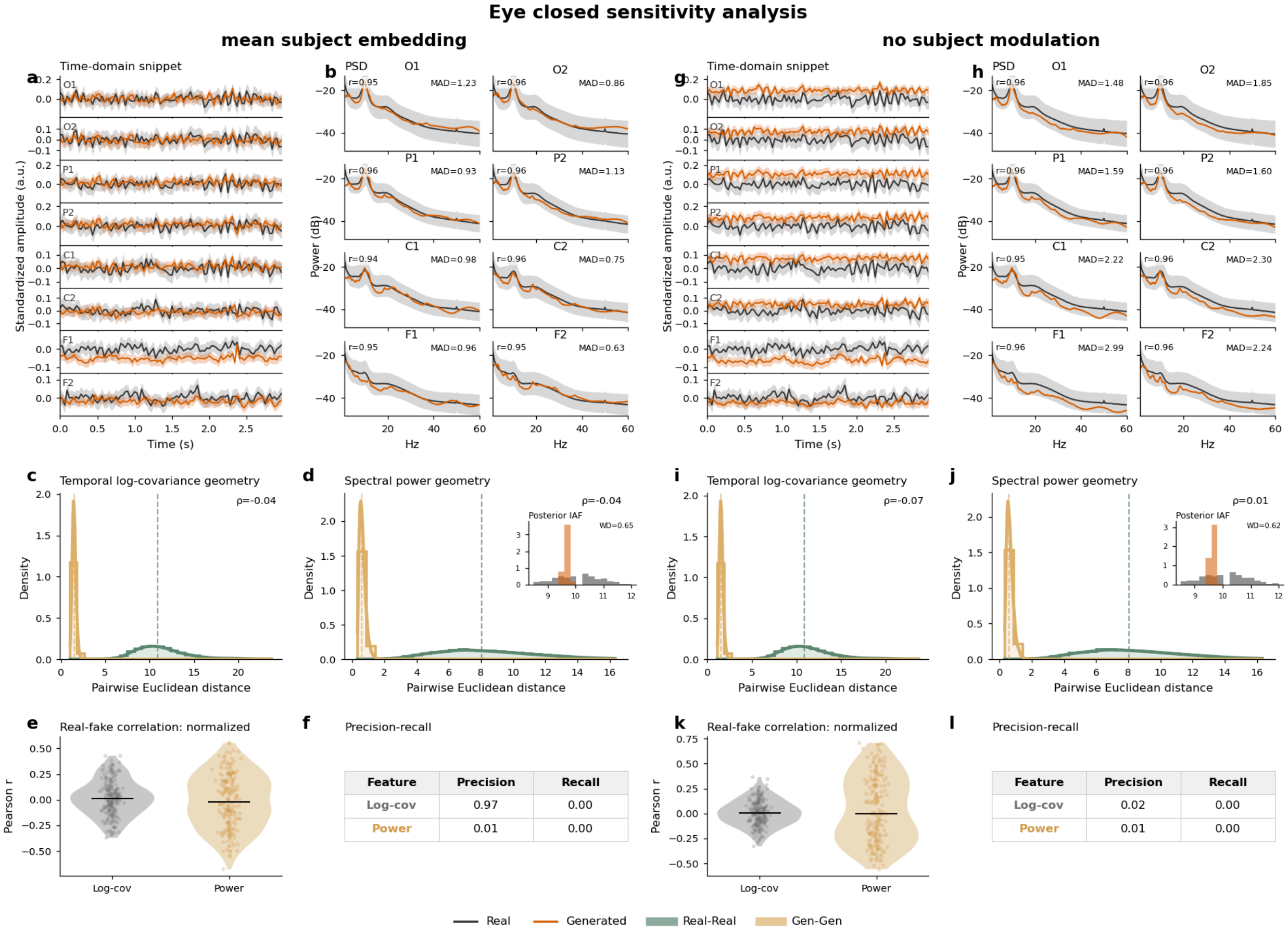

f Precision-recall

| Feature | Precision | Recall |
|---|---|---|
| Log-cov | 0.97 | 0.00 |
| Power | 0.01 | 0.00 |

l Precision-recall

| Feature | Precision | Recall |
|---|---|---|
| Log-cov | 0.02 | 0.00 |
| Power | 0.01 | 0.00 |

***Figure S1. Eye-closed sensitivity analysis of subject conditioning.*** *Comparison of generated EC signals using the* ***mean subject embedding*** *(a–f) or* ***no subject modulation*** *(g–l). Panels show representative time-domain snippets (a,g), channel-wise PSDs (b,h), real–real versus generated–generated geometry in log-covariance and spectral feature spaces (c,d,i,j), subject-wise real–generated similarity based on normalized data (e,k), and precision–recall summaries (f,l). Insets in d and j show posterior individual alpha frequency (IAF) distributions averaged across O1, O2, P1, and P2 channels.*

These results, however, do not imply that generation is strictly confined to the training samples. Because FiLM defines a continuous subject-embedding space, the model can also support out-of-distribution generation by interpolating between learned subject embeddings, thereby producing signals that were not seen during training. Figure S2 illustrates this possibility using power spectral density (PSD) profiles.

For the in-distribution examples shown in Figure S2a–b, the generator receives only latent noise together with the subject identity used to retrieve the corresponding learned embedding. In other words, at inference the model does not have any access to the target subject's real EEG segment, but instead generates a sample conditioned only on noise and the learned subject representation. For the out-of-distribution example in Figure S2c, we interpolated between the embeddings of the two displayed subjects and used the resulting intermediate embedding to generate a new synthetic subject. The metric labeled inside provides a simple channel-wise summary of how often the interpolated PSD falls between the PSDs of the two real subjects. Higher values indicate that the interpolated generation remains mostly within the spectral range spanned by the two subjects, whereas lower values suggest either extrapolation beyond that range or weaker agreement.

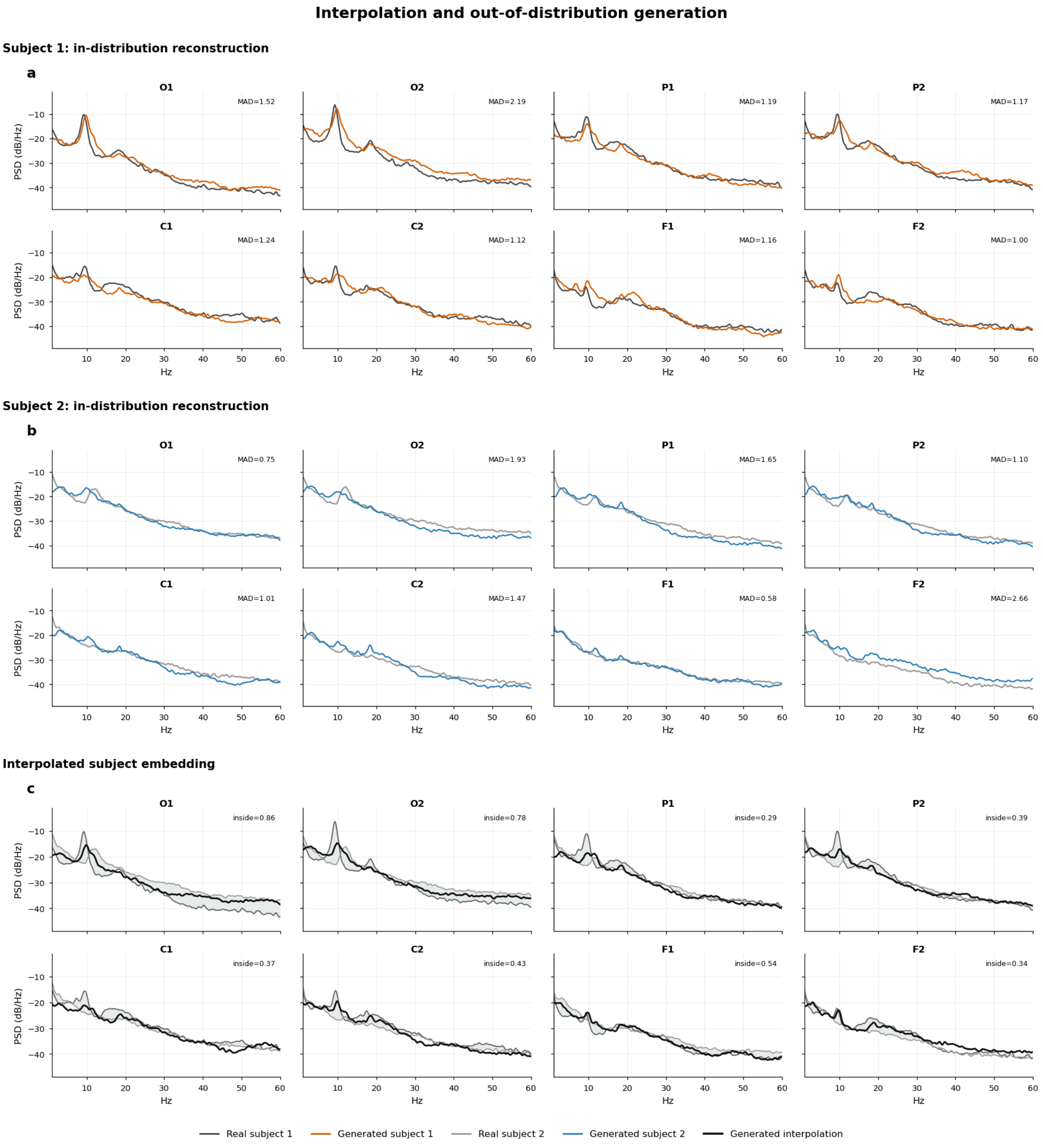

***Figure S2. Examples of in-distribution and out-of-distribution generation in the FiLM-conditioned model.*** *Panels a–b show PSDs of generated samples for two training subjects, demonstrating subject-specific reconstruction from latent noise and the learned subject embedding alone. Panel c shows generation from an interpolated subject embedding between these two subjects, illustrating out-of-distribution synthesis within the continuous embedding space. The inside value summarizes how often the interpolated PSD lies between the PSDs of the two real subjects across channels.*

## S2: Low-dimensional embedding analysis of real and generated EEG

We used Uniform Manifold Approximation and Projection (UMAP) to visualize the relationship between real and generated EEG segments in a two-dimensional embedding space. Each EEG segment was first represented by its flattened time-domain signal (512×8), and UMAP was applied to project these high-dimensional vectors into two dimensions. To complement the global projection, we selected representative anchor points from different regions of the UMAP space and identified their nearest opposite-class match in the original high-dimensional segment space. The corresponding time-domain traces are shown in Figure S3b. This allows visual inspection of real–generated segments similarity beyond the two-dimensional projection. Samples selected from shared or transition regions show broadly similar waveform structure, whereas samples from edge regions show more visible differences. The reported distance ***d*** denotes the Euclidean distance between flattened 512×8 segment vectors in the original high-dimensional space.

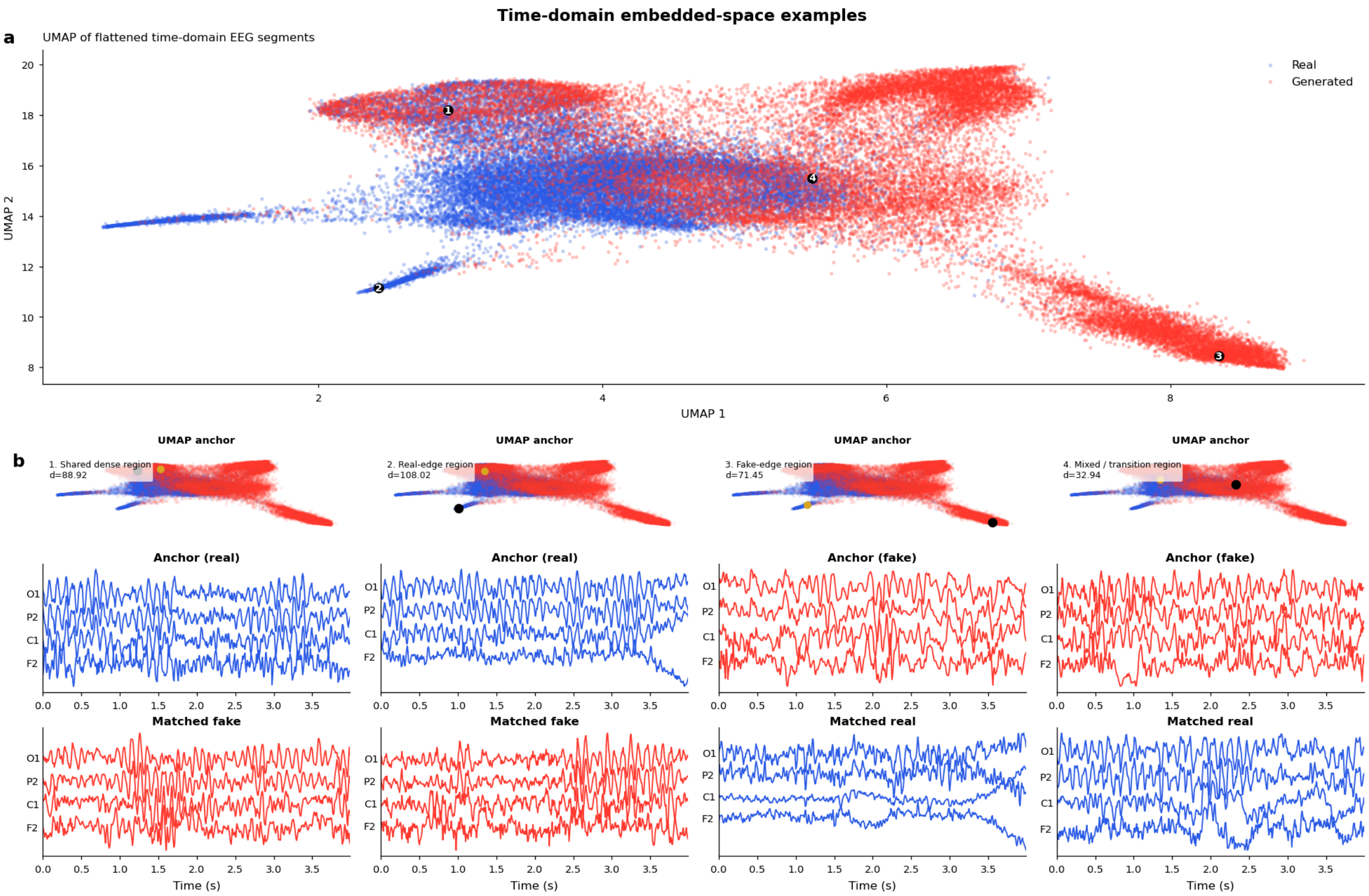

***Figure S3. UMAP visualization of real and generated EEG segments.*** *(a) Two-dimensional UMAP projection of flattened EEG segments, with real samples shown in blue and generated samples in red. Each point corresponds to one EEG segment. Numbered markers indicate the selected anchor regions. (b) Time-domain traces for each anchor segment and its nearest opposite-class match in the original high-dimensional segment space. These examples illustrate the correspondence between real and generated samples in shared, edge, and transition regions of the UMAP projection.*

## S3. Standardized Fréchet Distance trajectories and feature distributions for eye-open condition

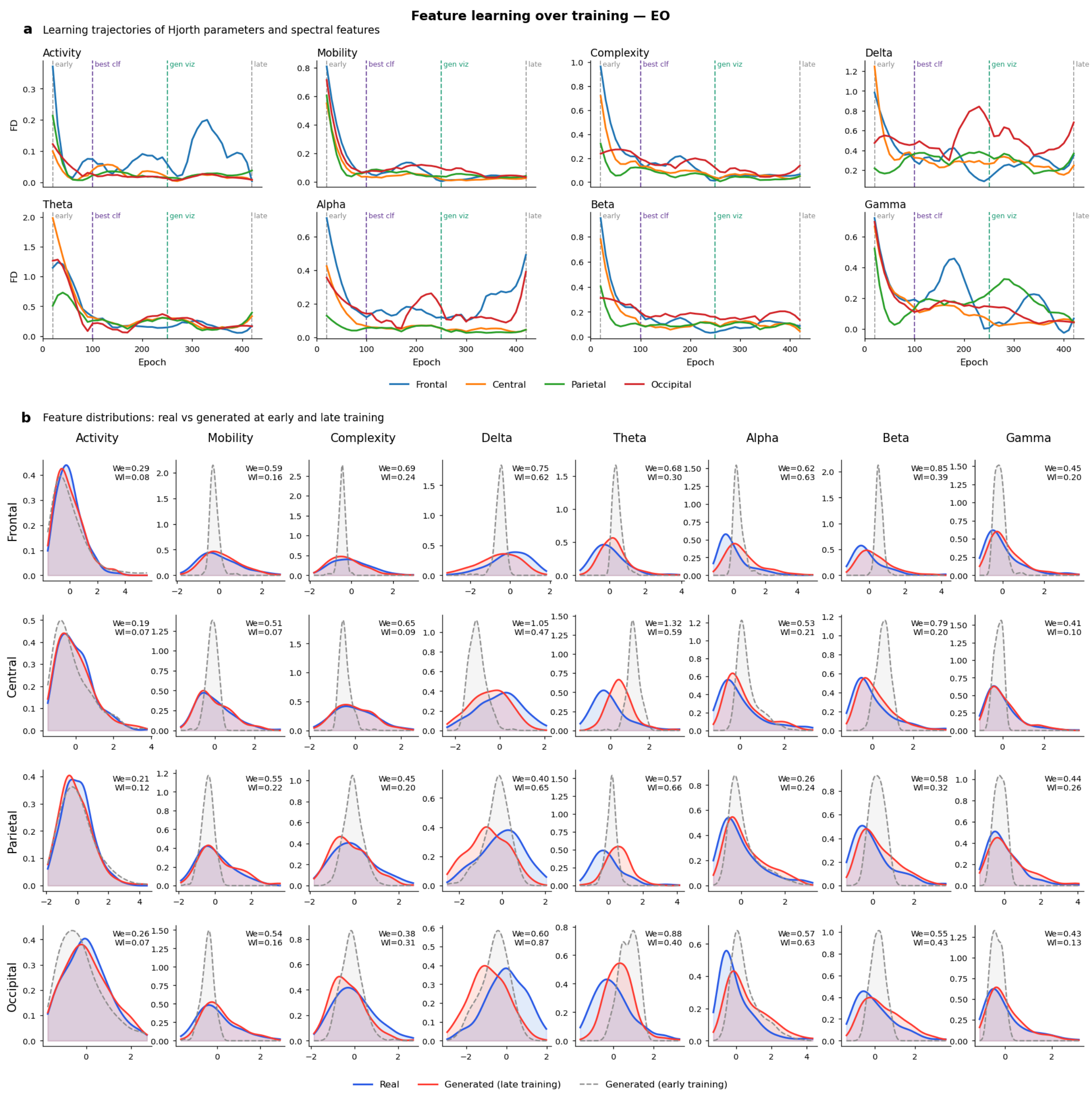

***Figure S4. Feature-distribution similarity during training in the EO condition. a,*** *Standardized Fréchet Distance (FD) trajectories for Hjorth and spectral-power features across frontal, central, parietal, and occipital channel groups. Vertical dashed lines mark selected training checkpoints.* ***b,*** *Real and generated feature distributions at early and late training for the same feature–region combinations. We and Wl denote Wasserstein distances between real and generated distributions at the early and late checkpoints, respectively.*

## S4. Model performance on hypnotizability data

To further test the utility of the extracted representations on a noisier and more variable label, we evaluated their performance on hypnotizability classification using an independent OpenNeuro dataset (Kekecs et al., 2023). In this dataset, participants' hypnotizability was measured with the Harvard Group Scale of Hypnotic Susceptibility (HGSHS), a self-report-based scale ranging from 0 to 12.

To frame this ordinal score as a binary classification task, we used a cut-off score of 5, corresponding to the median in the dataset. We then downsampled the data to obtain a roughly balanced number of participants in the high- and low-hypnotizability classes, resulting in 33 participants and 13,936 EEG segments. We compared the performance of features extracted from REST-GAN Critic, CBraMod, and flattened raw EEG segments. The results are summarized in Table S1.

These results are particularly interesting in light of our previous study (Farahzadi et al., 2024), where we used more extensive preprocessing, included all 56 EEG channels, and trained a support vector classifier on manually extracted features, including spectral power and connectivity measures. Even with this more elaborate pipeline, the highest accuracy obtained for the hypnotizability classification task was 63%. The present finding therefore further highlights the promise and reusability of the unsupervised features learned by our model, which appear to retain predictive information even for labels as noisy and subjective as self-assessed hypnotizability scores.

| **Dataset size** | **Feature source** | **Feature dimension** | **Val. accuracy** | **Val. balanced accuracy** | **Val. AUROC** |
|---|---|---|---|---|---|
| 33 subjects / 13,936 segments | Flattened raw EEG | 4096 | 0.506 ± 0.006 | 0.505 ± 0.005 | 0.509 ± 0.010 |
| 33 subjects / 13,936 segments | REST-GAN critic (ours) | **144** | **0.602 ± 0.111** | **0.607 ± 0.106** | **0.649 ± 0.15** |
| 33 subjects / 13,936 segments | CBraMod | 6400 | 0.556 ± 0.032 | 0.554 ± 0.038 | 0.586 ± 0.035 |

***Table S1.*** *Performance of different feature sources on the hypnotizability classification task. The highest performance for each metric is shown in bold.*